\documentclass{aastex62}

\usepackage{natbib}
\usepackage{inputenc}
\usepackage{amsmath,amsfonts}
\usepackage{bbold}
\usepackage{graphicx}
\usepackage{verbatim}
\usepackage{xspace}
\usepackage{bm}

\graphicspath{ {figs/} }

\newcommand{\phoebe}{PHOEBE\xspace}

\newcommand{\uvec}[1]{\bm{\hat{#1}}}

\newcommand{\ihat}{\bm{\hat{\textbf{\i}}}}
\newcommand{\jhat}{\bm{\hat{\textbf{\j}}}}
\newcommand{\khat}{\bm{\hat{\textbf{k}}}}
\newcommand{\uhat}{\bm{\hat{\textbf{u}}}}
\newcommand{\vhat}{\bm{\hat{\textbf{v}}}}
\newcommand{\what}{\bm{\hat{\textbf{w}}}}

\newcommand{\surfpot}{\widetilde\Omega}
\newcommand{\lobe}{\widetilde{\mathcal L}}

\renewcommand{\vec}[1]{\bm{#1}}

\shorttitle{Binary stars with spin-orbit misalignment revisited}

\shortauthors{Horvat et al.}

\begin{document}

\title{Physics of Eclipsing Binaries. III. Spin-Orbit Misalignment}

\author[0000-0002-0504-6003]{Martin Horvat}
\affiliation{Villanova University, Dept.~of Astrophysics and Planetary Sciences, 800 E Lancaster Ave, Villanova PA 19085, USA}
\affiliation{University of Ljubljana, Dept.~of Physics, Jadranska 19, SI-1000 Ljubljana, Slovenia}

\author[0000-0002-5442-8550]{Kyle E.~Conroy}
\affiliation{Villanova University, Dept.~of Astrophysics and Planetary Sciences, 800 E Lancaster Ave, Villanova PA 19085, USA}
\affiliation{Vanderbilt University, Dept.~of Physics and Astronomy, 6301 Stevenson Center Ln, Nashville TN, 37235, USA}

\author[0000-0002-1355-5860]{Herbert Pablo}
\affiliation{American Association of Variable Star Observers, 49 Bay State Road, Cambridge, MA 02138, USA}

\author{Kelly M.~Hambleton}
\affiliation{Villanova University, Dept.~of Astrophysics and Planetary Sciences, 800 E Lancaster Ave, Villanova PA 19085, USA}
\affiliation{University of Ljubljana, Dept.~of Physics, Jadranska 19, SI-1000 Ljubljana, Slovenia}

\author[0000-0002-9739-8371]{Angela Kochoska}
\affiliation{Villanova University, Dept.~of Astrophysics and Planetary Sciences, 800 E Lancaster Ave, Villanova PA 19085, USA}

\author{Joseph Giammarco}
\affiliation{Eastern University, Dept.~of Astronomy and Physics, 1300 Eagle Rd, St.~Davids, PA 19087}

\author[0000-0002-1913-0281]{Andrej Pr\v sa}
\affiliation{Villanova University, Dept.~of Astrophysics and Planetary Sciences, 800 E Lancaster Ave, Villanova PA 19085, USA}
\affiliation{University of Ljubljana, Dept.~of Physics, Jadranska 19, SI-1000 Ljubljana, Slovenia}

\email{martin.horvat@fmf.uni-lj.si}

\begin{abstract}
Binary systems where the axis of rotation (spin) of one or both components is tilted w.r.t.~the axis of revolution are called \emph{misaligned} binary systems. The angle of misalignment, obliquity, has been measured for a handful of stars and extrasolar planets to date. Here we present a mathematical framework for a complete and rigorous treatment of misalignment and introduce an extension to the public PHOEBE code that implements this framework. We discuss misalignment for the Roche geometry and introduce methods for computing stellar shapes, equilibrium (generalized Lagrange) points of the potential and minimal requirements for lobe existence. Efficient parameterization of misalignment is proposed in the plane-of-sky coordinates and implementation details in PHOEBE are given alongside the proof-of-concept toy model, comparison with a known misaligned binary DI Her, and comparison with a misaligned planetary system Kepler-13. We provide important mathematical details of the model in the Appendix. This paper accompanies the release of PHOEBE 2.1, which is available from its website \texttt{http://phoebe-project.org}.
\end{abstract}


\keywords{line: profiles, binaries (including multiple): close, binaries: eclipsing, binaries: spectroscopic, stars: fundamental parameters}

\section{Introduction}

At first glance it is very tempting to think that stars in stellar systems would have their rotational axes aligned with the orbital axis: the total angular momentum during a protostellar cloud contraction is conserved, so we expect a high degree of retained symmetry. Yet this is not what we observe in nature.

Looking at our Solar System alone, we see that misalignment abounds.  The angle between the axes of rotation and revolution (or, conversely, between the equatorial and orbital planes) is called \emph{axial tilt} or \emph{obliquity}. If obliquity is $0$, then the two axes are aligned. Starting with the Sun, its equator is tilted to the ecliptic by $7.25^\circ$. Earth's equator is tilted on average $23.44^\circ$ and it precesses at a rate of $\sim50''$/year, so obliquity changes as a function of time. Solar system planets all orbit very close to the ecliptic plane (within $\sim7^\circ$), but their rotational axes are nowhere near aligned, ranging from $0.2^\circ$ to $82^\circ$ \citep{aa2018}. 

Transiting exoplanet host stars exhibit a wide range in their obliquities, from completely aligned (i.e.~HD 189733; \citealt{winn2006}), to moderately aligned (i.e.~XO-3 at $37^\circ$; \citealt{hirano2011}), to perpendicular (WASP-7; \citealt{albrecht2012}) and even retrograde (WASP-17; \citealt{anderson2010}). Giant extrasolar planets on very close, eccentric orbits (the so-called hot and warm Jupiters) also demonstrate a large range and oscillatory behavior in their obliquities \citep{dawson2014}. Two good reviews on the methodology and results of obliquity measurements in exoplanet systems are done by \citet{winn2015} and \citet{triaud2017}. 

Lastly, there are two shining examples for misalignment among eclipsing binary stars: DI Her \citep{albrecht2009, philippov2013} and CV Vel \citep{albrecht2014}. DI Her is the current record holder with the sky-projected spin-orbit angles of $\beta_p = 72^\circ \pm 4^\circ$ and $\beta_s = -84^\circ \pm 8^\circ$ for the primary and secondary star, respectively, while CV Vel features a misaligned primary at $\beta_p = -52^\circ \pm 6^\circ$ and a (sky-projected) aligned secondary at $\beta_s = 3^\circ \pm 7^\circ$.

The most commonly used method to measure misalignment is to acquire spectroscopic observations of a binary star during the eclipse. The measured radial velocity is a weighted average of individual radial velocities across the visible elements of the star. In a misaligned system, the transiting star no longer passes along the parallel of the eclipsed star, resulting in an asymmetry in the Rossiter-McLaughlin effect (RME; \citealt{rossiter1924, mclaughlin1924}). Furthermore, as the obliquity affects the distortion of the eclipsed star, it will also affect the intensities and weighting of the individual eclipsed elements (most notably due to gravity brightening). Together, these deviations from the aligned case are called the \emph{anomalous} Rossiter-McLaughlin effect.

The analytical formalism for binary systems with misaligned rotational and orbital angular velocity vectors has been discussed previously by \cite{limber1963, kruszewski1966, avni1982}. It was not until now, however, that this effect has been built into an eclipsing binary (EB) modeling code. A few recent reports \citep{winn2006,albrecht2012,triaud2013, harding2013} use the anomalous RME to measure obliquity, but none of these provide any technical insight into their treatment of \emph{tidally and gravitationally distorted} binary systems. Our goal is to provide that here, along with a publicly available tool to model misaligned cases. We discuss different aspects of modeling binary systems with spin-orbit misalignment. We focus on properly defining the model and its parameters in Section 2 and discuss equilibrium (generalized Lagrange) points of the misaligned potential in Section 3. In Section 4, we demonstrate the misalignment treatment with a toy model as implemented in the open-source software package \phoebe \citep{prsa2016a}, and on two observed systems: DI Her and Kepler-13. In the Appendix, we provide further technical details about the model, along with mathematical tools to compute poles, area,s and volumes that are used for a robust synthesis of observables (light curves, radial velocity curves, and spectral line profiles). 

\section{Effective potential} \label{sec:eff_pot}

In this section we introduce the framework for misaligned spin axes in binaries. We use the following nomenclature: vectors are denoted with boldface (i.e.~$\vec{r}$); vectors with unit magnitude (unit vectors) are denoted with a \ $\bm{\hat{}}$ \ symbol (i.e.~$\uvec{r}$); vector magnitudes (norms) are slanted (i.e.~$r \equiv \| \vec r \|$); and fractional, unitless values are denoted with greek letters (i.e.~$\rho = r/a$) with $a$ being the semi-major axis of the considered binary system.

\begin{figure}[t]
\centering
\includegraphics[width=0.8\textwidth]{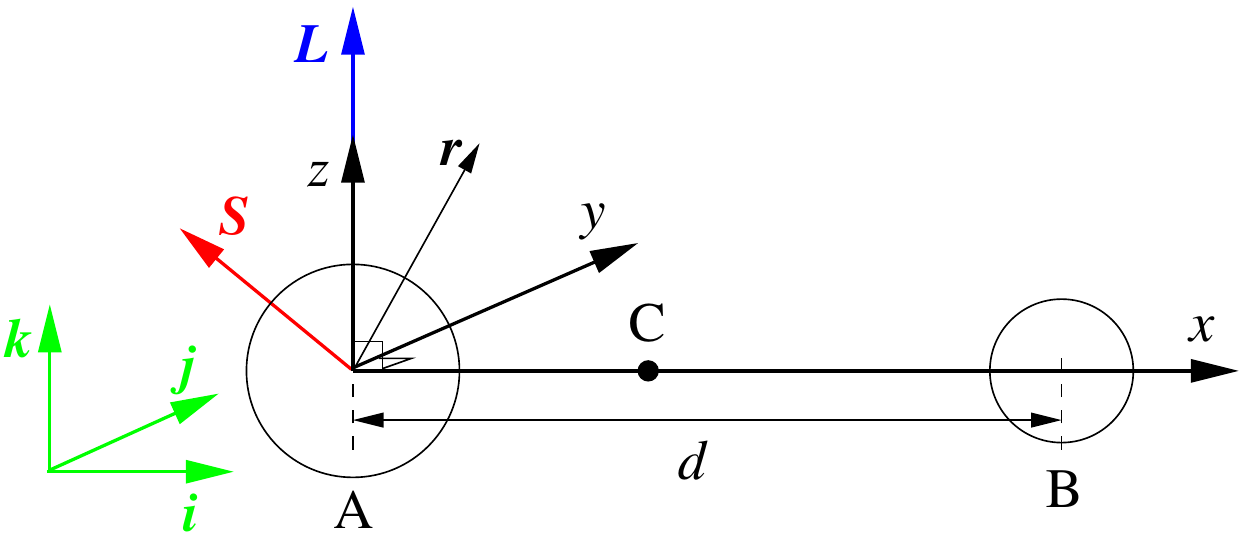}
\caption{Schematic diagram of a binary system comprised of stars A and B. The center of mass is at point C. The canonical coordinate system, $(x,y,z)$, has the origin at and is co-rotating with the orbital motion of star A. The orbital angular momentum, denoted by $\vec L$, is aligned with the $z$-axis. The axis of rotation of star A is denoted by $\uvec S$. The distance between the centers of both stars is $d$.}
\label{fig:binary}
\end{figure}

Let two stars (labeled A and B) rotate in the $xy$ plane of the inertial coordinate system about the common center of the mass (labeled C) with an angular frequency $\omega_{\rm L}$. The angular momentum $\vec L$ of the binary system points in the direction of the $z$-axis of the inertial coordinate frame. Let star A rotate uniformly with angular velocity $\vec{\omega}_\mathrm{S} = \omega_\mathrm{S} \uvec{S}$ about its center of mass, where $\uvec{S}$ is the unit spin vector. We introduce a \emph{canonical} coordinate system with the origin at the center of star A, the $x$-axis pointing toward the center of star B and the $z$-axis aligned with orbital axis (i.e.~the $z$-axis of the inertial system). We denote the orthogonal vector basis of the canonical coordinate system by $(\ihat, \jhat, \khat)$. A schematic of the binary system and the canonical coordinate system are depicted in Fig.~\ref{fig:binary}.

The lobes of the stars are defined as the surfaces of constant pressure and density. These lobes are approximated by iso-surfaces of the potential $V$, which is written in the canonical coordinate system as:
\begin{equation}
    V(\vec r) = 
    -\frac{GM_\mathrm{A}}{r}
    -\frac{GM_\mathrm{B}}{\|\vec r - d \ihat\|}
    +\frac{GM_\mathrm{B}}{d^2} (\ihat \cdot \vec{r})
    -\frac{1}{2} \omega_{\mathrm S}^2 \|({\uvec S} \cdot \vec{r}){\uvec S} - \vec{r} \|^2 \>,
    \label{eq:kopal_potential}
\end{equation}
with $\vec{r}$ denoting the position and  $d$ denoting the distance between the stars. The masses of stars A and B are labeled by $M_\mathrm{A}$ and $M_\mathrm{B}$, respectively. A detailed discussion and derivation of the potential $V$ for a circular orbit can be found in \citet{limber1963} and its generalization to the non-circular case, as used here, is presented in \citet{avni1982} and is summarized in Appendix \ref{sec:deriv_kopal}. 
Eq.~(\ref{eq:kopal_potential}) can be further simplified by introducing a dimensionless potential $\Omega(\vec{\rho})$, where $\vec{\rho} = \vec{r}/a$:
\begin{equation} \label{eq:nodimpot}
    V(a {\vec \rho}) =  
    -\frac{GM_\mathrm{A}}{a} \Omega(\vec \rho; q, F, \delta, \uvec{S})
\end{equation}
where:
\begin{equation}
    \Omega(\vec \rho; q, F, \delta, \uvec S) 
    = \frac{1}{\rho} + 
    q \left(\frac{1}{\| \vec{\rho} - \delta\, \ihat\|} - \frac{\ihat \cdot \vec{\rho}}{\delta^2} \right) 
            + \frac{1}{2} (1 + q) F^2 \| (\uvec S \cdot \vec{\rho}) \uvec S - \vec{\rho} \|^2.
    \label{eq:poten_orig}
  \end{equation} 
This form (for the aligned, circular, synchronous case) was first proposed by \citet{kopal1978} and generalized to eccentric, asynchronous orbits by \citet{wilson1979}. Here $q=M_\mathrm B/M_\mathrm A$ is the mass ratio, $F = \omega_{\mathrm S}/\omega_{\mathrm L}$ is synchronicity parameter, and $\delta \equiv d/a$ is fractional instantaneous separation. We use Kepler's third law to replace $\omega_{\rm L}^2 = (GM_{\mathrm A} + GM_{\mathrm B})/a^3$. Note that the Kopal potential is invariant to the sign of the vector $\uvec S$: $\Omega|_{ -{\uvec S} } = \Omega|_{\uvec S}$.

Next we introduce a rotated coordinate system about the $x$-axis w.r.t.~the canonical coordinate system so that vector $\uvec S$ lies in the new $xz$ plane. The vector basis $(\ihat', \jhat', \khat')$ of the rotated coordinate system is related to the canonical vector basis $(\ihat, \jhat, \khat)$ by the following relations:   
\begin{equation}
  \ihat' =  \ihat,\qquad
  \jhat' =  \jhat \cos\alpha + \khat \sin\alpha,\qquad
  \khat' = -\jhat \sin\alpha + \khat \cos\alpha,
  \label{eq:rot_basis}
\end{equation}
where $\alpha = -\arctan (\uvec S \cdot \jhat / \uvec S \cdot \khat)$. The positions are denoted by $\vec{r}' = x' \ihat' + y' \jhat' + z' \khat'$. Consequently, the vector $\uvec{S}$ can be given as
\begin{equation}
    \uvec{S} = \sin\beta\, \ihat' + \cos\beta\, \khat' \>,
\end{equation}
with the angle $\beta = {\rm arcsin}(\uvec{S}\cdot \ihat') \in [-\pi/2,\pi/2]$. We can concentrate on $\beta\ge 0$ without any loss of generality because negative $\beta$ correspond to the mirroring across the $xy$ plane ($z \mapsto -z$), cf.~Appendix \ref{sec:symm}.

\begin{figure}[t]
\begin{center}
\includegraphics[width=0.59\textwidth]{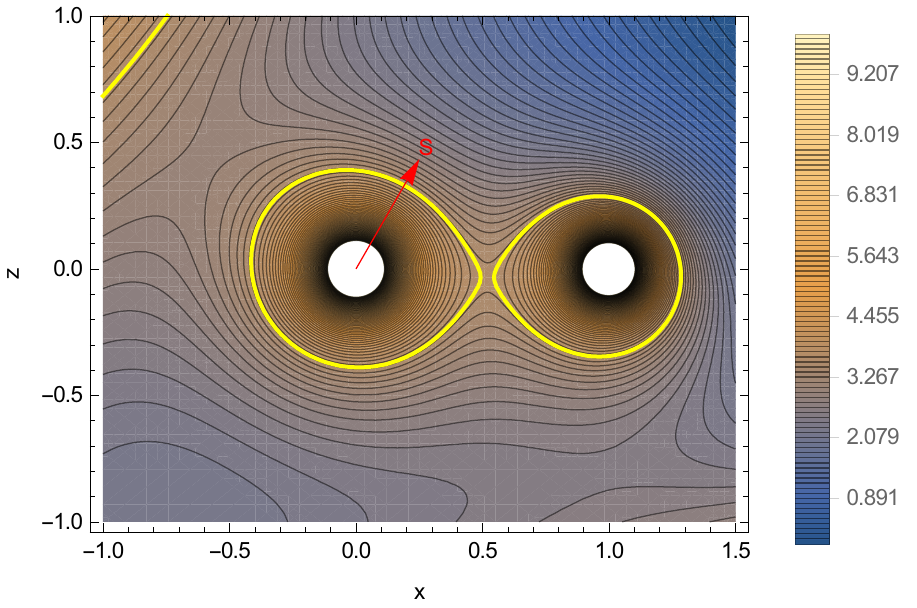}
\includegraphics[width=0.39\textwidth]{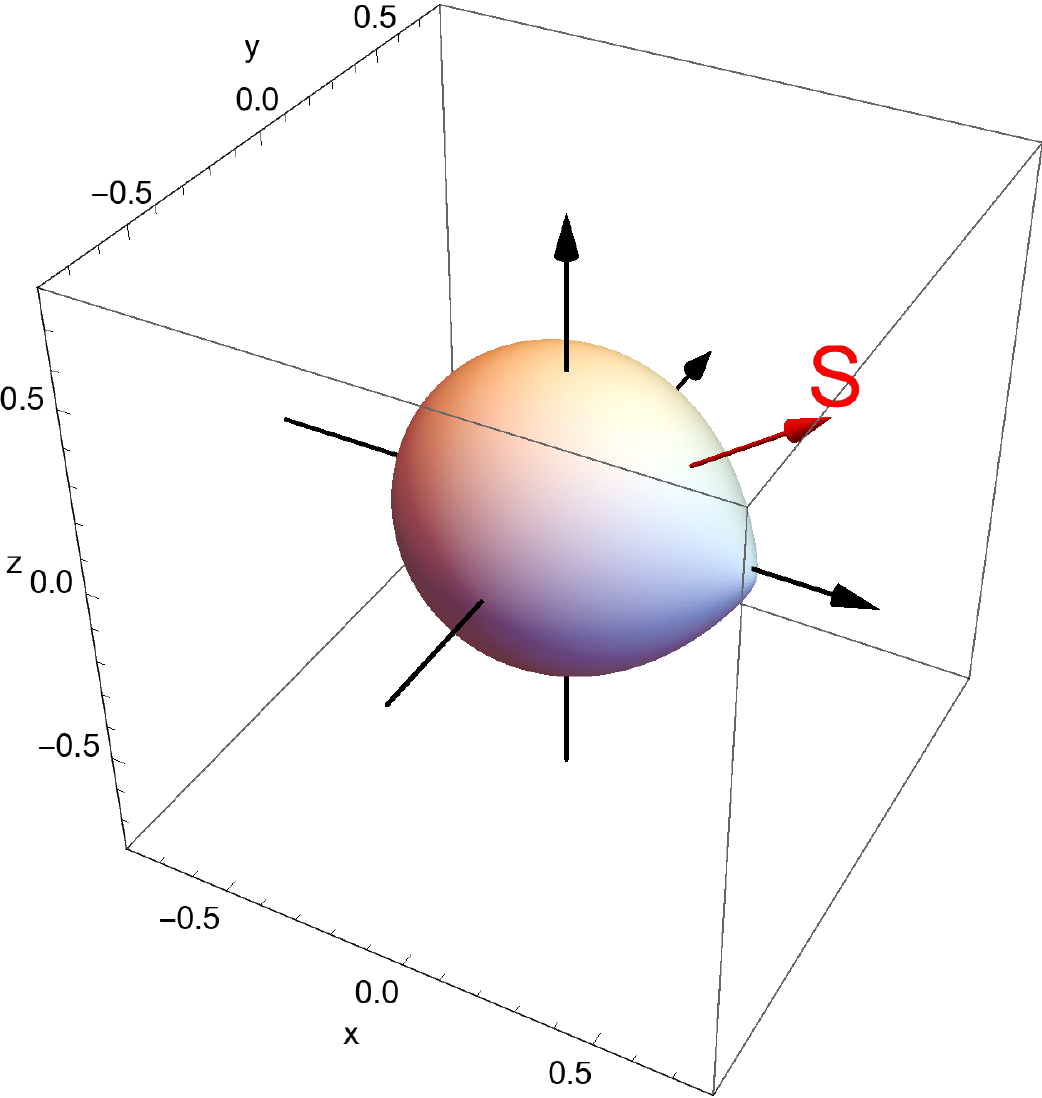} \\
\caption{Misaligned equipotential $\surfpot$ for $b=2$, $\beta = 0.3\pi$ and $q=1$. Left: $\surfpot$ contours; a contour at $\surfpot=3.6$ is drawn in yellow. Right: a lobe that corresponds to $\surfpot = 3.6$.}
\label{fig:roche_lobe}
\end{center}
\end{figure}

We present our analysis on the rescaled Kopal potential $\surfpot$ where $\rho' \mapsto \rho' \delta$ and $\Omega \mapsto \Omega \delta$, defined in the rotated coordinate system as
\begin{align}
   \surfpot (\vec{\rho}'; q, b, \beta) 
    &= \Omega (\vec{\rho}' \delta; q, F, \delta, \uvec{S}) \delta \nonumber\\
    &= \frac{1}{\rho'} + 
    q \left(\frac{1}{\| \vec{\rho}' - \ihat'\|} - \ihat' \cdot \vec{\rho}'\right) 
    + \frac{1}{2} b\left[ (x' \cos \beta - z' \sin\beta)^2 + {y'}^2\right]\>,
   \label{eq:poten_red}    
\end{align}
where $b = (1 + q)F^2 \delta^3$ is an auxiliary parameter. The shape of the star is fully determined by the value of $\surfpot$: it corresponds to the iso-surface of the potential. We refer to this shape as the \emph{lobe} and denote it with $\lobe$. Fig.~\ref{fig:roche_lobe} depicts the $\surfpot$ contours that correspond to $q=1$, $b=2$ ($F = 1$, $\delta = 1$) and $\beta = 0.3\pi$ (left) and the lobe that corresponds to $\surfpot = 3.6$ (right). 

\section{Equilibrium points of the Kopal potential}

Equilibrium points of the Kopal potential $\Omega$ (Eq.\,\ref{eq:poten_orig}) are defined as the points where the gradient of the potential equals 0.  The simplified case of the aligned ($\beta=0$), synchronous ($F=1$) equipotential has been extensively studied. The corresponding equilibrium points are called Lagrange points. There are 5 such points: $L_1$, $L_2$ and $L_3$ lie on the $x$-axis, while $L_4$ and $L_5$ lie in the $xy$ plane, forming an equilateral triangle with the two massive bodies. $L_4$ and $L_5$ can thus be computed analytically, while $L_1$, $L_2$ and $L_3$ are computed\footnote{Good analytic approximations exist for $L_1$, $L_2$ and $L_3$, given, i.e., by \citet{taff1985}.} numerically. For a recent study of the analytic properties of Lagrange points see \cite{seidov2004}.

In the aligned and non-synchronous ($F\neq 1$) case there are still five equilibrium points $\vec{G}_i (q,b)$, $i=1,\ldots,5$ \citep{kallrath2009}, which are generalizations of the Lagrange points. In the $F \to 1$ limit, these points are identical to Lagrange points. For the purposes of binary star physics, the first three generalized Lagrange points are of most interest. Obtaining their values for arbitrary $q$ and $b$ using generic algorithms for solving nonlinear equations can be time-consuming and, in some cases, unstable. Because of this, we developed a specialized numerical solver to stabilize and speed up the process, which we have implemented into \phoebe 2.0 \citep{prsa2016}. The solver is based on novel analytical approximations of generalized Lagrange points in difference regimes of parameters $(q,b)$ that are further polished via the Newton-Raphson scheme or with the Laguerre method \citep{sirca2012}.

For the reduced potential $\surfpot$ (Eq.~\ref{eq:poten_red}), the condition $\nabla \surfpot(\vec{K}_i)=0$ that determines the equilibrium points $\vec{K}_i = (x',y',z')$ can be written as
\begin{equation} \label{eq:poten_grad}
  \left ( 
    b \left [
    \begin{array}{ccc}
      \cos^2\beta & 0 & -\sin\beta \cos\beta \\
      0 & 1 & 0 \\
      -\sin\beta \cos\beta & 0 & \sin^2\beta
    \end{array}
    \right]
    - \left (
    \frac{1}{r_1^3} + \frac{q}{r_2^3}
  \right) {\bf id}
  \right) \vec{K}_i + 
  q \left ( 
    \frac{1}{r_2^3} - 1
  \right)\ihat = 0 \>,
\end{equation}
with $b = (1+q)F^2 \delta^3$, $r_1 \equiv \rho' = \| \vec{K}_i\|$ and $r_2 = \| \vec{K}_i - \ihat\|$. The equilibrium points are crucial for understanding the global behavior of the potential and for determining the necessary condition for lobe existence. The solutions of Eq.~(\ref{eq:poten_grad}) can be generally divided into two groups:

\begin{itemize}

\item[(i)] The points outside the $xz$ plane: For $b - q \ge (\sqrt{2} |\cos\beta|)^{-3}$, we find that there are two equilibrium points of the reduced Kopal potential $\surfpot$:
\begin{equation}
    \left(
        \xi, 
        \pm \sqrt{2\xi - \frac{\xi^2}{\cos^2\beta}}, 
        -\xi \tan \beta
    \right)\>,
\end{equation}
with $\xi = \frac{1}{2} [b - q]^{-\frac{2}{3}}$, which is a further generalization of the Lagrange points $\vec{G}_{4,5}$. If, on the other hand, $b-q < (\sqrt{2} |\cos \beta |)^{-3}$, there are no real equilibrium points.  

\item[(b)] The points in the $xz$ plane. In the aligned case ($\beta=0$), there are exactly three equilibrium points $\vec{G}_i(q,b)$ for $i=1,2,3$ and all are saddle points on the $x$-axis. In the misaligned case ($\beta\neq 0$), however, there can be more than three points and they can be of a different type (minimum, maximum or a saddle point).
\end{itemize}

The points in the latter group determine the smallest value of the potential for which the lobe of the primary star exists. We discuss this group of equilibrium points next.

\subsection{Phenomenology} 

\begin{figure}[t]
\centering
\includegraphics[width=15cm]{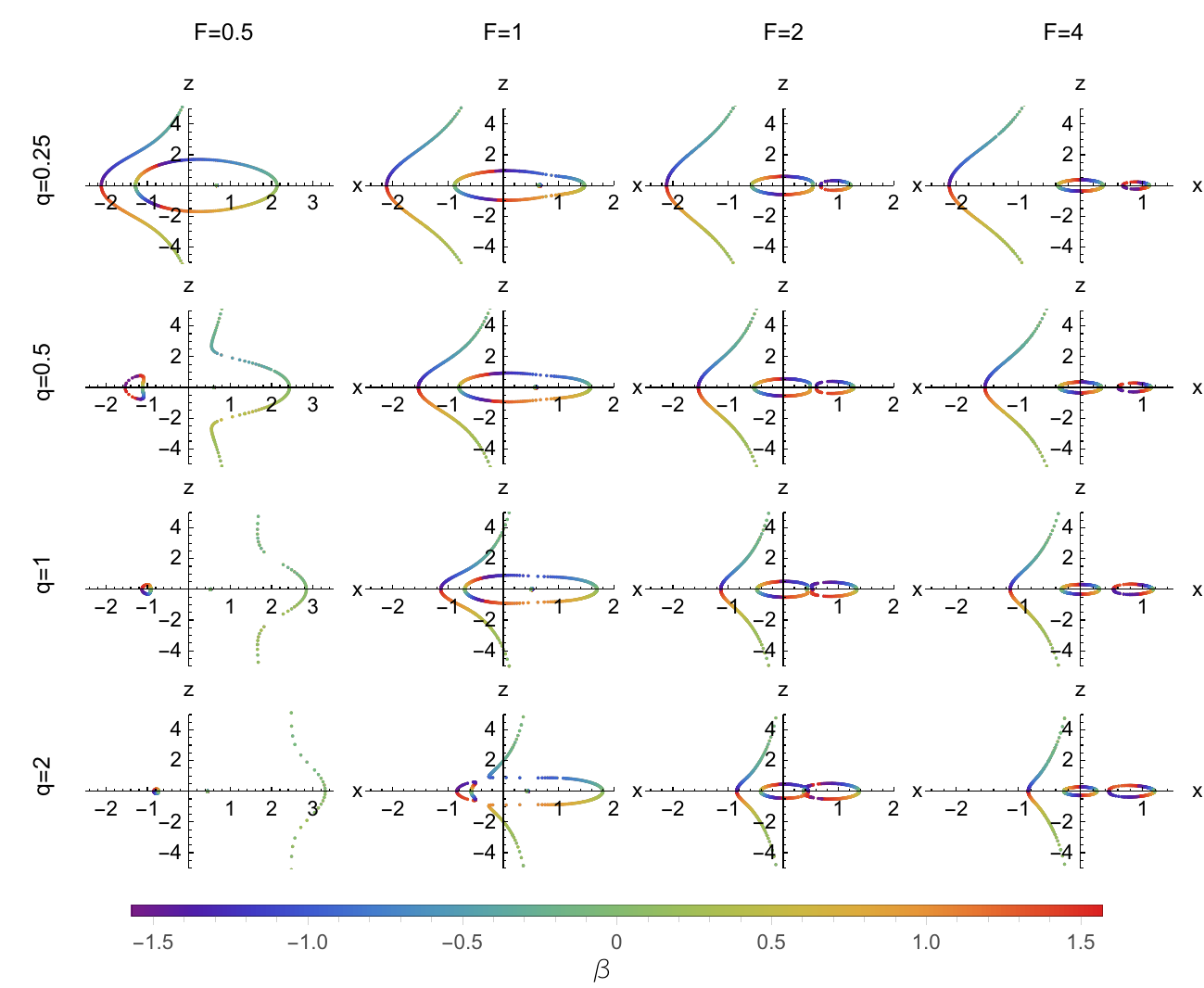}
\caption{Equilibrium points of the reduced Kopal potential $\surfpot (x, 0, z; q,  b, \beta)$ at a given $q$ and $b$, as a function of $\beta \in [-\pi/2,\pi/2]$. The color of the points corresponds to the value of $\beta$.}
\label{fig:fps2}
\end{figure}

Finding the equilibrium points of the reduced potential $\surfpot$ (Eq.\,\ref{eq:poten_red}) in the $xz$ plane is a non-trivial computational task. Fig.~\ref{fig:fps2} depicts the equilibrium points as a function of the misalignment parameter $\beta$, $\beta \in [-\pi/2,\pi/2]$ and several values of $q$ and $b$. The color of the points corresponds to the value of $\beta$. We see that the position of the equilibrium points varies continuously with $\beta$ over a large range of values, but there are some discontinuities represented by the lack of points. Note the symmetry in the positions of points across the $x$-axis. 

\begin{figure}[t]
\centering
\includegraphics[width=15cm]{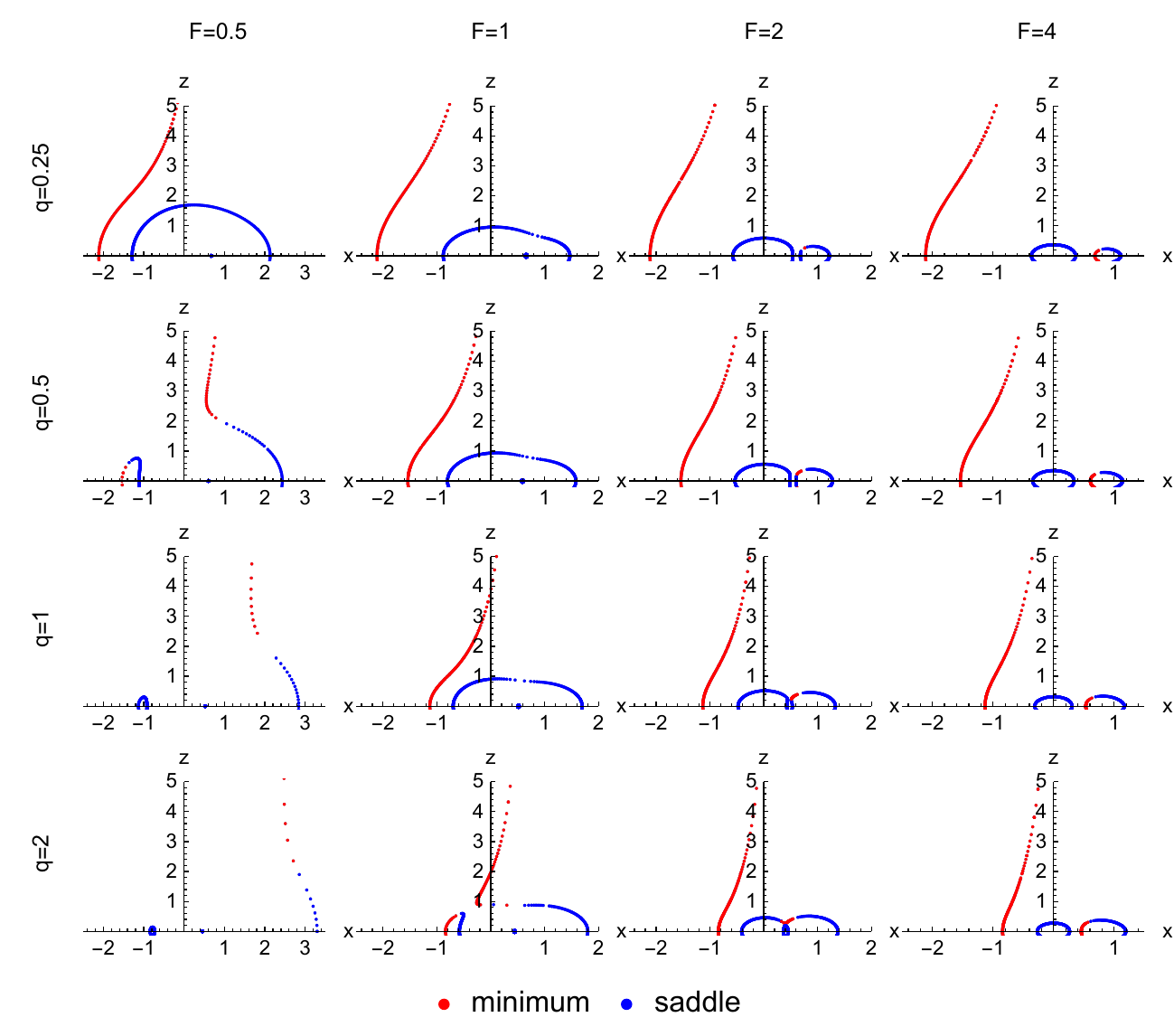}
\caption{Equilibrium points of the reduced Kopal potential $\surfpot (x, 0, z; q,  b, \beta)$ for the positive $z$-axis at a given $q$ and $b$. Blue color corresponds to saddle points $\vec{s} = (-1,1), (1,-1)$ and red color corresponds to local minima $\vec{s} =(1,1)$. There are no local maxima.}
\label{fig:fps1}
\end{figure}

We distinguish three types of equilibrium points in the $xz$ plane: saddle points, local minima, and local maxima. Let $\{h_k(\vec{r}):k=1,2\}$ be the eigenvalues of the Hessian matrix $(\nabla_{xz} \otimes \nabla_{xz}) \surfpot (\vec{r})$. These are proportional to the local principal curvatures, and the type of an equilibrium point is determined by the sign of the principal curvatures, $\mathrm{sign}(h_k(\vec{K}_i))$ \citep{carmo2016}. The types of equilibrium points are important for drawing qualitative conclusions about the shape of the nearby isosurfaces. Fig.~\ref{fig:fps1} depicts the same equilibrium points as Fig.~\ref{fig:fps2}, but here the colors denote their type. The saddle points are of most interest as they determine the separatrix, i.e.~they yield the limiting value of the potential for which the lobes exist. If we consider equilibrium points of any given type as ``branches'', we see that, by perturbing a certain parameter, branches can cross, meaning that parts of the branches change their type. 

\subsection{A method for finding equilibrium points}

We developed and present here an efficient method to obtain a subset of equilibrium points in the $xz$ plane.  The location of the equilibrium points that correspond to the misaligned potential is found by tracing the variation in the $x$-location of the equilibrium points that correspond to the aligned potential as the misalignment parameter $\beta$ is varied. This procedure is significantly faster than the general nonlinear root finding algorithm employed in constructing Figs.~\ref{fig:fps2} and \ref{fig:fps1}. The method is applicable for the values of $\beta = 0$ up to the value $\beta_{C,i}$ for the $i$th equilibrium point at which the Hessian becomes singular.

Let us denote with $\mathcal{I}$ an interval around the aligned value $\beta = 0$ at which the location of the equilibrium point $\vec K_i$ varies smoothly with $\beta$ at any constant $q$ and $b$. We can then write:
\begin{equation}    
    \nabla \surfpot(\vec{K}_i(\beta); q, b, \beta) = 0  
    \qquad \forall \beta \in {\cal I} \>,
    \label{eq:crit_curve}
\end{equation}
where $\vec K_i$ is the $i$th generalized Lagrange point at $\beta=0$:
$$
\vec{K}_i(\beta = 0) = \vec{G}_i(q,b).
$$
By differentiating Eq.~(\ref{eq:crit_curve}) w.r.t.~$\beta$, we obtain a differential equation that determines the equilibrium point manifold as a function of $\beta$:
\begin{equation}
  \frac{{\rm d}}{{\rm d} \beta }\vec{K}_i(\beta) =
 - \left[(\nabla\otimes \nabla)  \surfpot(\vec{K}_i(\beta); q, b, \beta)\right]^{-1}
   \cdot 
   \nabla \frac{\partial}{\partial \beta}\surfpot(\vec{K}_i(\beta); q, b, \beta) \>. 
  \label{eq:rc_evol}
\end{equation}
To obtain the $i$th equilibrium point for a given $(q,b,\beta_0)$, we integrate Eq.~(\ref{eq:rc_evol}) over the range $[0,\beta_0]$ with the initial condition $\vec{K}_i(0) = \vec{G}_i(q,b)$. This can be done numerically by an ordinary differential equation integrator, e.g.~a 4th order Runge-Kutta \citep{sirca2012}, as long as the Hessian $(\nabla \otimes \nabla) \surfpot$ is non-singular. The Hessian is for all three Lagrange points non-singular on the entire $\beta_0 \in [-\pi/2, \pi/2]$ range only for $q=F=\delta=1$; for other parameter values, $\beta_0 \in [-\beta_{C,i}, \beta_{C,i}]$ for the $i$th equilibrium point. Typically, the smallest is $\beta_{C,3}$. Thus, a singular Hessian pinpoints the transition between the types of equilibrium points, as depicted in Fig.\,\ref{fig:fps1}.

\subsection{Minimal value of the Kopal potential for the existence of lobes}

\begin{figure}[t]
\centering
\includegraphics[width=12cm]{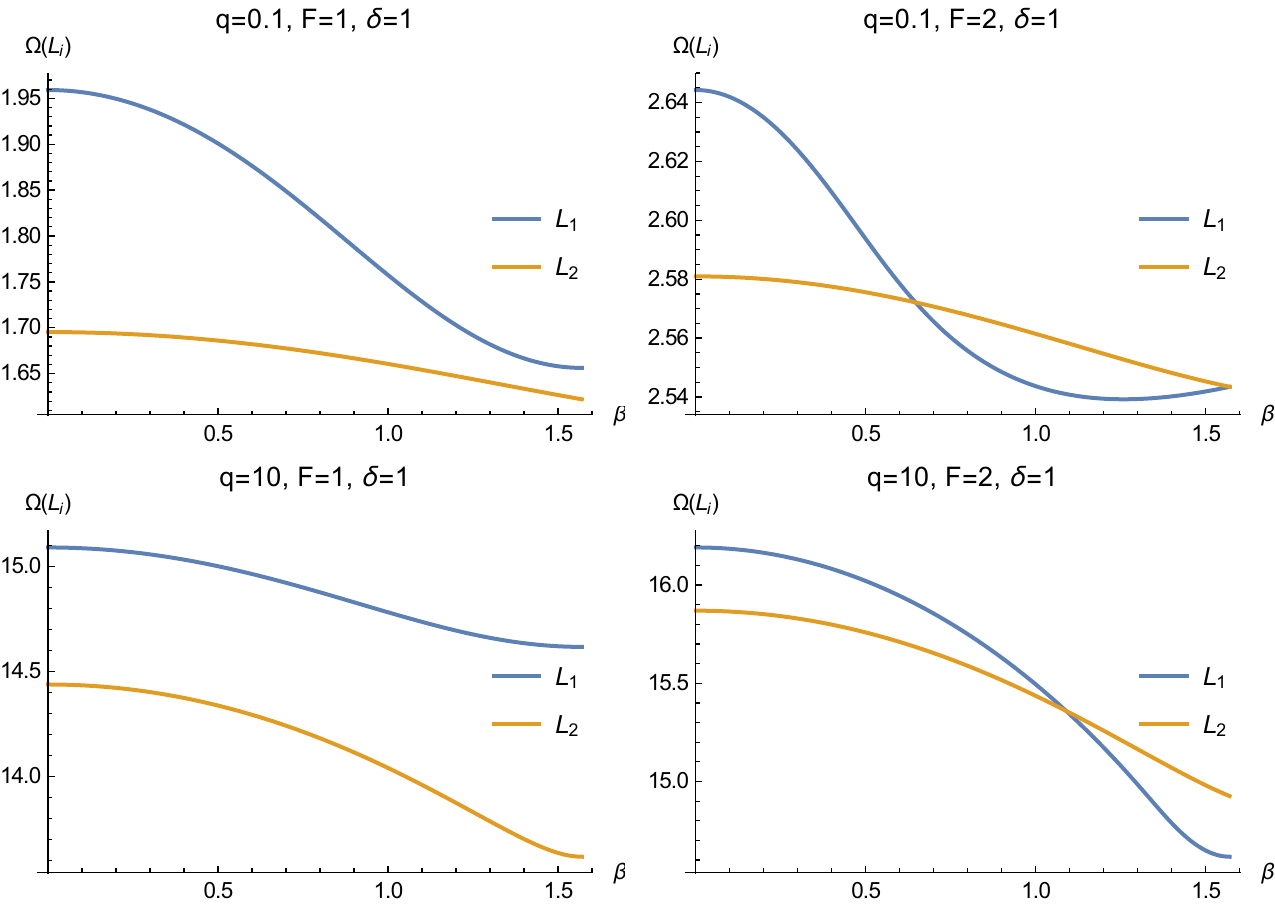}
\caption{The value of the reduced potential $\widetilde\Omega$ (Eq.\,\ref{eq:poten_red}) in equilibrium points $\vec{K}_1(q,b,\beta)$ and $\vec{K}_2(q, b,\beta)$ as a function of the misalignment angle $\beta$ for various values of parameters $q$ and $b$ with $\delta=1$.}
\label{fig:omega_min}
\end{figure}

With $\vec{K}_i(q, b; \beta)$ being the misaligned generalizations of their aligned counterparts $G_i(q, b)$, the existence of a detached primary star lobe is determined by the values of $\vec{K}_1(q, b; \beta)$ and $\vec{K}_2(q, b; \beta)$. If the potential is smaller than either value, the equipotential will not delimit a closed surface. The minimal value of the reduced Kopal potential $\surfpot_{\rm min}$ for which a detached primary lobe exists is thus the maximal value of the potentials at equilibrium points $\vec{K}_1$ and $\vec{K}_2$: 
\begin{equation}
  \surfpot_{\rm min} (q,b,\beta) =
  \max 
    \left\{
        \surfpot \left(\vec{K}_1(q,b;\beta)\right),  
        \surfpot \left(\vec{K}_2(q,b;\beta)\right)
    \right\} \>.
  \label{eq:omega_min}
\end{equation}
The lobe at $\surfpot_{\rm min} (q,b,\beta)$ thus represents a generalized Roche lobe. Fig.\,\ref{fig:omega_min} demonstrates how the value of the potential changes with $\beta$ for $\vec{K}_1(\beta; q, b)$ and $\vec{K}_2(\beta; q, b)$. As already pointed out by \cite{avni1982}, the curves of the potential values associated with both equilibrium points can intersect,  meaning that the roles of the equilibrium points in constraining the lobes can change as $\beta$ changes. The angle of intersection, referred to as the \emph{critical angle}, was closely analyzed by \citeauthor{avni1982}. The minimum value of the potential for which a lobe exists at a certain angle $\beta$ is the maximum value of both of these curves. 

\section{Orbital misalignment in the plane-of-sky}

\begin{table}[t!]
\caption{
\label{tab:params}
Principal parameters of the toy model of a misaligned binary star.}
\begin{center}
\begin{tabular}{lccc}
  \hline \hline
  parameter: & & system: & \\
            & primary star: & & secondary star: \\ 
  \hline
  semi-major axis $a[R_\odot]$ & & 3.98 & \\
  period $P[{\rm d}]$ & & 0.65 & \\
  mass ratio $q$ & & 0.7 & \\
  eccentricity $\epsilon$ & & 0 & \\
  inclination $i [^\circ]$ & & 80 & \\
  long. of ascending node $\Omega[^\circ]$ & & 0 & \\
  systemic velocity $\gamma[{\rm km/s}]$ & & 0 & \\
  \hline
  atmosphere & blackbody & & blackbody \\
  equivalent radius $R[R_\odot]$ & 1.3 & & 0.8\\
  effective temperature $T_{\rm eff}[K]$ & 6500 & & 5500\\
  synchronicity parameter $F$ & 1 & & 3.61 \\
  \hline
  LD model & logarithmic & & logarithmic \\
  LD coefficient $x_{\rm LD}$ & 0.5 & & 0.5 \\ 
  LD coefficient $y_{\rm LD}$ & 0.5 & & 0.5 \\
  gravity darkening $\beta_\mathrm{grav}$ & 0.32 & & 0.32 \\
  \hline
  $\Delta \Omega[{}^\circ]$ & 0 & & 45 \\
  $\Delta i [{}^\circ]$ & 0 & & 45 \\
  \hline
\end{tabular}
\end{center}
\end{table}

The orbit of a binary system is described in the canonical coordinate system, where the orbital plane coincides with the $xy$ plane. The plane-of-sky is the plane perpendicular to the line-of-sight. The corresponding coordinate system, spun by unit vectors $\uhat$, $\vhat$ and $\what$, is oriented so that $\uhat$ and $\vhat$ lie in the plane-of-sky and point toward east and north, respectively, and $\what$ points toward the observer. To place the orbit in space w.r.t.~the observer, we use three angles: longitude of the ascending node $\Omega$, argument of periastron $\omega$ and inclination $i$.  The transformation from the orbital plane to the plane-of-sky is given by the following rotation:
$$
    \vec R_{uw} (i,\Omega) = \vec R_w(\Omega) \vec R_u(-i),
$$
where $\vec R_w$ and $\vec R_u$ are rotation matrices about the $w$ and $u$ axes, respectively. The definitions of rotation matrices here in use are given in Appendix \ref{sec:rot_mat}. The direction of the angular momentum $\uvec{L}$ is then given by
$$
  \uvec{L} = \vec R_{uw} (i, \Omega) \what = [-\sin i \sin \Omega, \sin i \cos \Omega, \cos i ]^T \>.
$$
Likewise, the spin vector $\uvec{S}$ can be written as
$$
  \uvec{S} = {\bf R}_{uw}(i', \Omega') \hat {\bf w} = [-\sin i' \sin \Omega', \sin i' \cos \Omega', \cos i' ]^T \>,
$$
where $i'$ and $\Omega'$ denote the inclination and longitude of the ascending node w.r.t.~the rotated coordinate system, and are related to the orbital inclination $i$ and longitude of the ascending node $\Omega$ by
$$
  \Omega' = \Omega + \Delta \Omega,\qquad  i' = i + \Delta i \>.
$$
%

The angle differences $\Delta \Omega$ and $\Delta i$ uniquely describe spin misalignment. The spin vector $\uvec{S}$ in the canonical coordinate system can then be written as
$$
  \uvec{S} = {\vec R}_w (-\varpi){\vec R}_u(i) {\vec R}_w (\Delta \Omega) {\vec R}_u(-i - \Delta i) \what,
$$
where $\varpi = \omega + \omega_0 + \nu$ is argument o latitude of the considered star and is a sum of the true anomaly $\nu$, argument of periastron $\omega$ and its positional offset $\omega_0$, which depends on the star: $\omega_0=\pi$ for the primary star and $\omega_0=0$ for the secondary star.

\subsection{Implementation in \phoebe}

\phoebe is an open source modeling suite developed for the analysis of single, binary and multiple stellar systems. Its initial version, released in 2005 and described in \citet{prsa2005}, was built on top of the widely used \citet{wilson1971} code and it was specifically designed for the modeling of eclipsing binary stars (hence the name, PHysics Of Eclipsing BinariEs). The updated version, PHOEBE 2.0, was released in 2016 and described by \citet{prsa2016}. It constitutes a complete rewrite that generalizes the algorithms to single and multiple stellar systems. This work further expands the functionality of PHOEBE 2 by the implementation of misalignment described in the sections above and in the Appendices, and is accompanied by the PHOEBE 2.1 release. It is available at \texttt{http://phoebe-project.org}.

\begin{figure}[t!]
\centering
\includegraphics[width=0.6\textwidth]{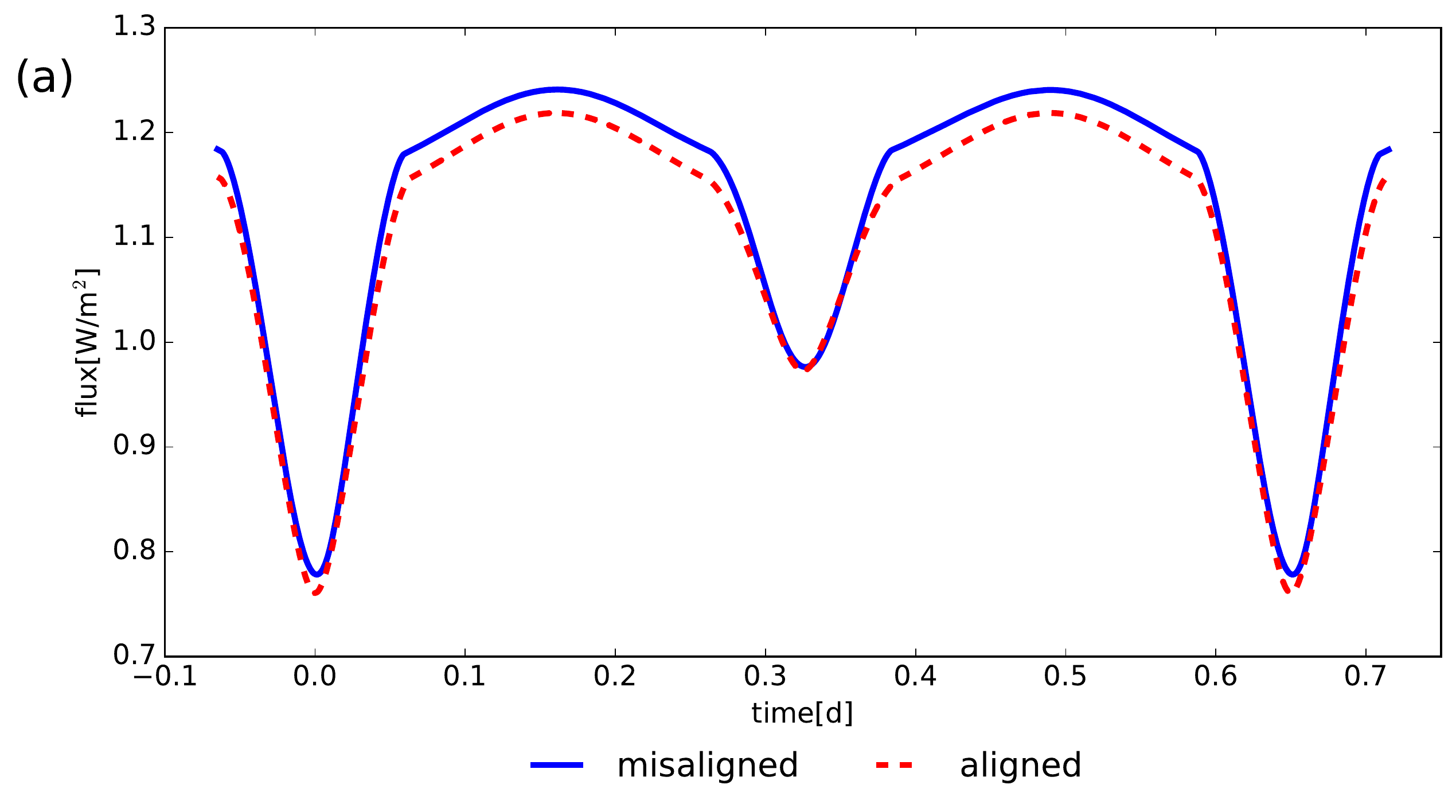} \\
\includegraphics[width=0.6\textwidth]{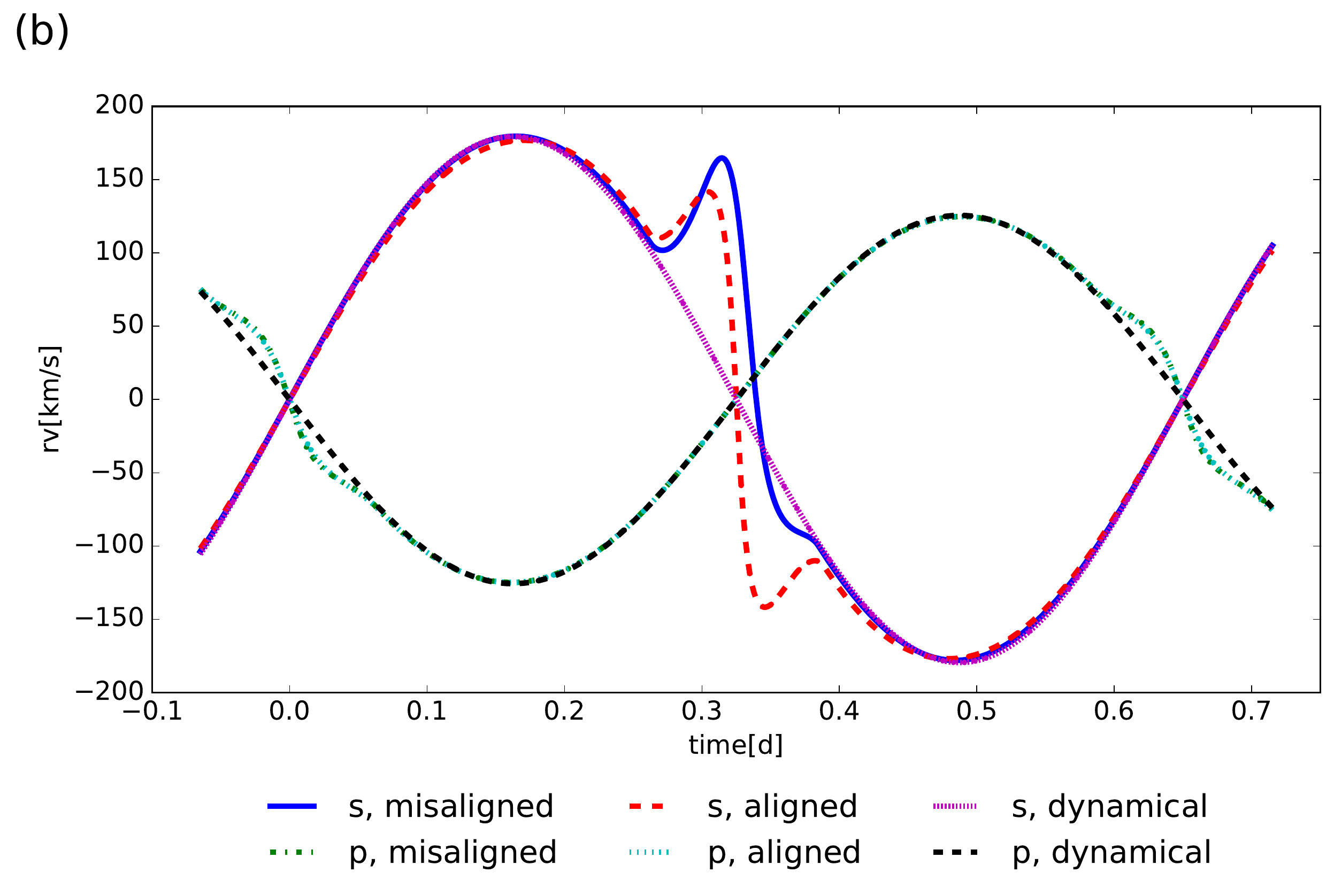}
\caption{Comparison of the light curves (a) and radial velocity curves (b) for an aligned and a misaligned binary system with parameters given in Table \ref{tab:params}. Designations $p$ and $s$ stand for primary and secondary star, respectively. As reference, the radial velocity plot depicts dynamical radial velocity curves for both stars. The effect of misalignment is clearly pronounced in both types of observables.}
\label{fig:lc_rv}
\end{figure}

PHOEBE 2.1 introduces spectral line profiles as a new type of dataset. The line profiles are computed from fiducial spectral lines (i.e. a Gaussian or a Lorentzian profile) at the user-provided rest wavelength, Doppler-shifted at each local surface element and weighted by the passband brightness distribution across the visible surfaces. Line profiles are provided in normalized flux units and do not include any slopes due to continuum or passband effects.

\begin{figure}[t!]
\centering
\includegraphics[width=0.495\textwidth]{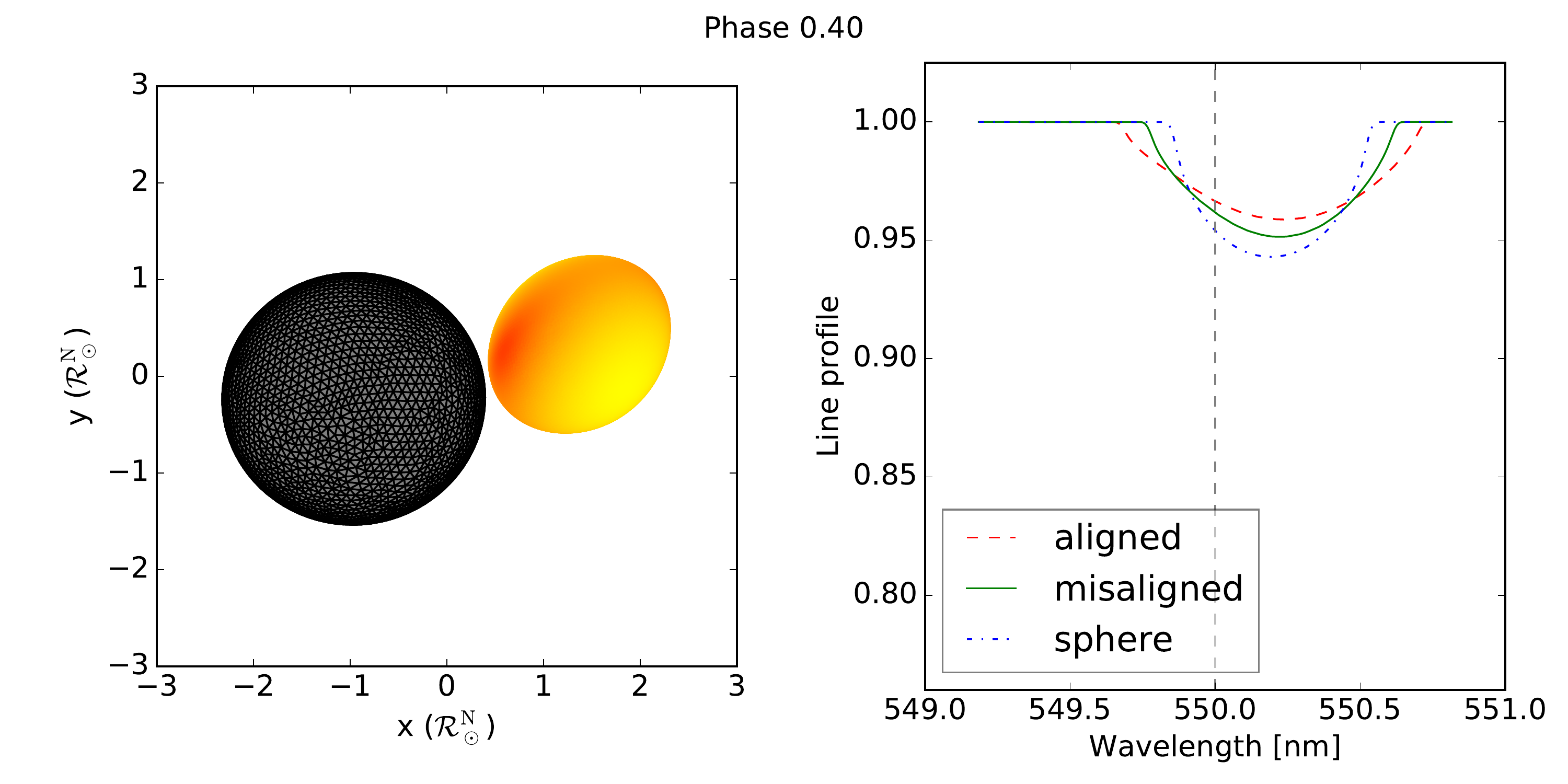}
\includegraphics[width=0.495\textwidth]{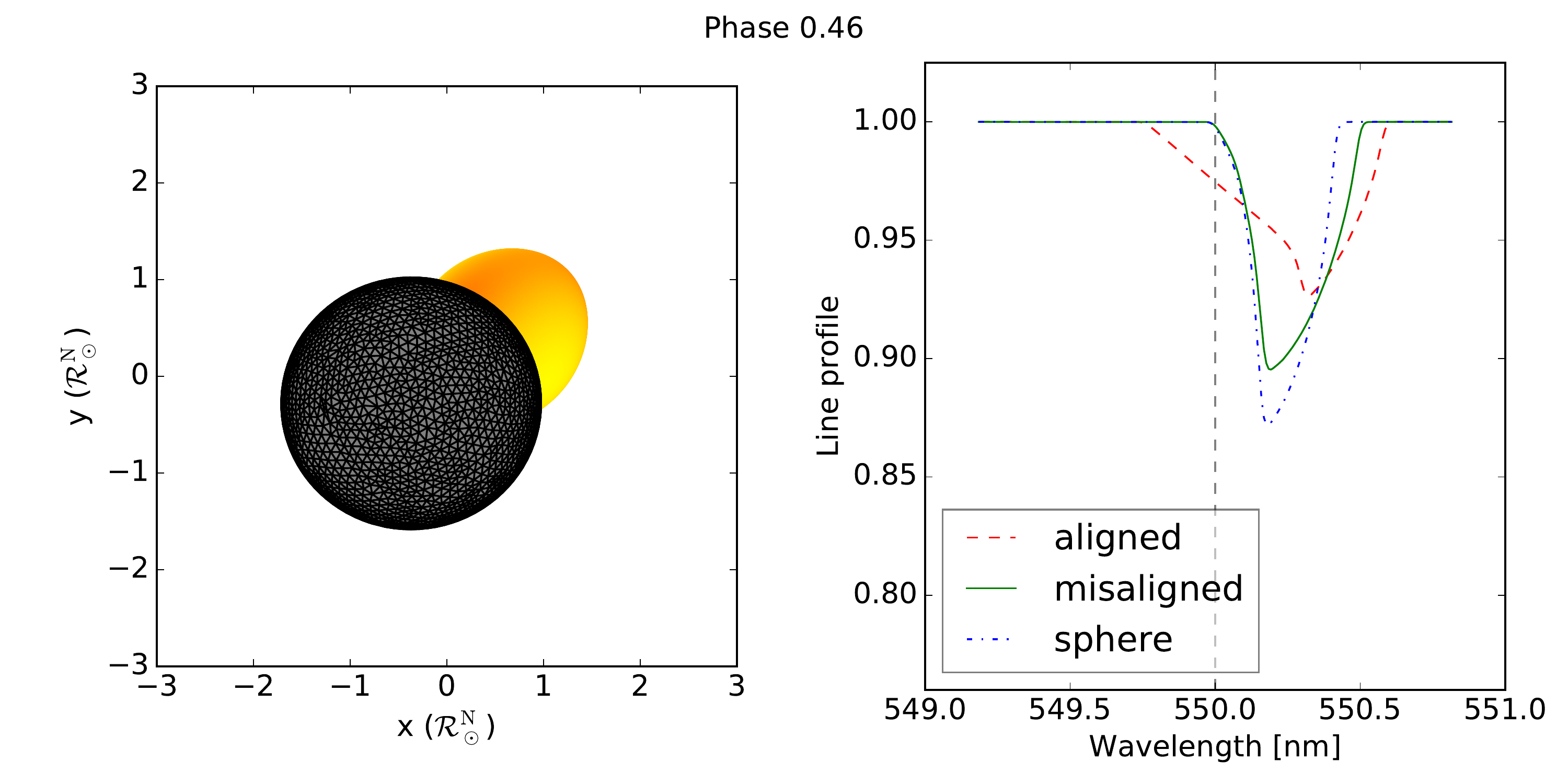} \\
\includegraphics[width=0.495\textwidth]{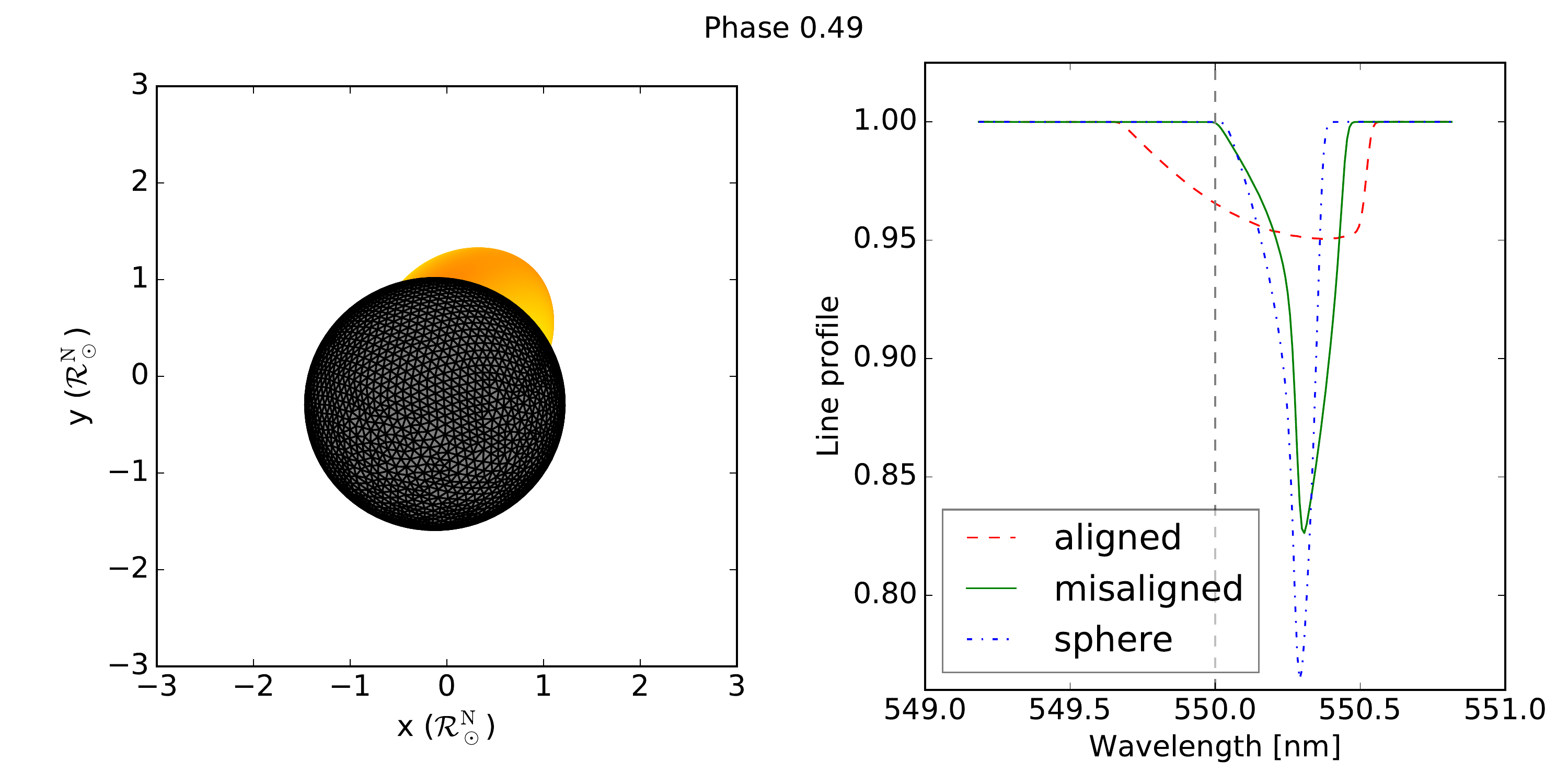}
\includegraphics[width=0.495\textwidth]{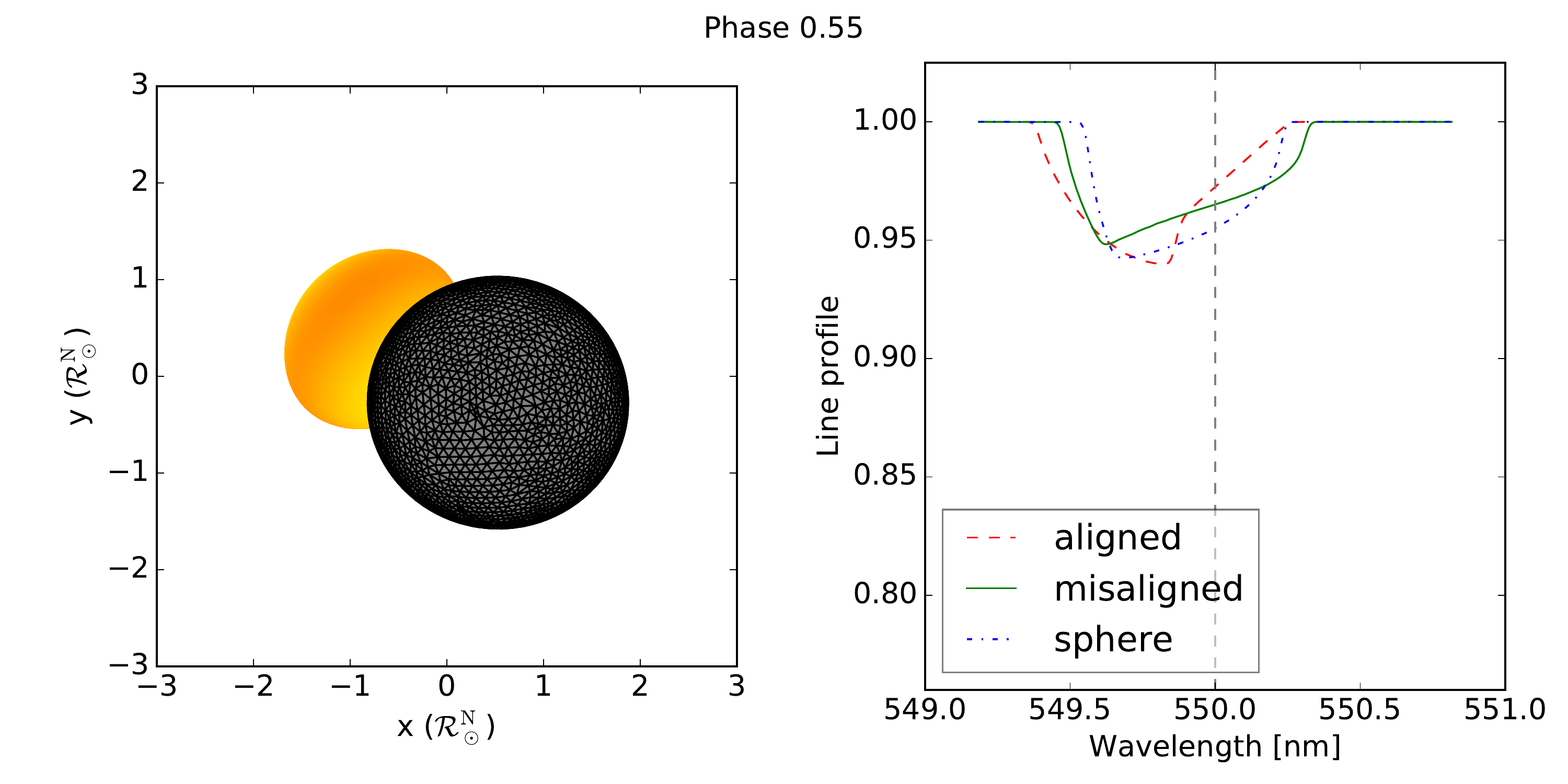} \\
\caption{Comparison of the position of lobes projected to plane of sky (left panel) and line profiles (right panel) at phases $\phi = 0.40$ and $0.46$ (top), and $0.49$ and $0.55$ (bottom). The lobe of the primary star is represented by its mesh used in computing and the lobe of the secondary star is color-coded by the surface temperature. The dashed gray line denotes rest wavelength of the line; the solid green, the dashed red and the dashed-doted blue lines represent the line profiles of the misaligned, the aligned systems and the spherical star system, respectively.}
\label{fig:lp}
\end{figure}

Using PHOEBE, we demonstrate the effect of misalignment on astrophysical observables (light curves, radial velocity curves and spectral line profiles) for a toy-model binary system with a misaligned secondary star. The parameters of the toy model are given in Table \ref{tab:params}. Fig.~\ref{fig:lc_rv} showcases the comparison between the misaligned system and the aligned system with the matching equivalent radius (the radius of the sphere that has the same volume as the bounding equipotential) of each component. Spin misalignment clearly has a significant effect on all observables. A telltale sign of misalignment is an asymmetry in light curves, although asymmetries can arise from other physical effects as well, such as ellipsoidal variation and reflection in eccentric systems, spots, etc. Radial velocity curves are similar, with the telltale difference obvious in the eclipses (the Rossiter-McLaughlin effect). Fig.~\ref{fig:lp} depicts line profiles for several phases in the aligned system (dashed line), the misaligned system (solid line) and the spherical system (dashed-dotted line) with matching equivalent radii. As expected, the differences are the largest during eclipses, as the main driver for the line profile is the sum of the local intensities weighted by the projected surface element area. Even outside the eclipses, though, the widths of the line profiles can be substantially different (top left panel) because of the modified surface brightness distribution across the disk of the secondary star. Other effects, such as relativistic gravitational redshift \citep{takeda2012}, convective blueshift \citep{shporer2011}, micro- and macroturbulence \citep{steffen2013}, also affect the line profiles; while those can in principle be included in the computation within the PHOEBE framework, we did not include them in the simulation in order to quantify the influence of misalignment by itself.

%
%

Due to the selected misalignment parameters of the toy model, the corresponding light curve has a brighter out-of-eclipse region than the aligned model light curve. This is because the hotter polar regions are tilted toward the observer, thus contributing excess flux w.r.t.~the aligned case. The generally asymmetric excess flux is more prominent in distorted (i.e.~close) and/or rapidly rotating systems, where gravity darkening causes a significant variation of surface brightness across the stellar disk(s).

Radial velocity curves exhibit a well-known Rossiter-McLaughlin effect \citep{rossiter1924,mclaughlin1924}, which describes a deviation from the dynamical (i.e.~center-of-mass) radial velocity curve due to eclipses that block certain parts of the star and thus induce a bias in the photometrically weighted mean radial velocity curve for each component. The Rossiter-McLaughlin effect is symmetric for the aligned case, but is generally\footnote{We say generally because a obliquity of $\pm 90^\circ$ would also lead to a symmetric effect.} asymmetric in the misaligned case. This is depicted in Fig.~\ref{fig:lc_rv}: the effect is symmetrical near timestamp 0 where the aligned primary star is eclipsed, and asymmetrical near timestamp 0.32 where the misaligned secondary star is eclipsed.

In consequence, light and radial velocity curves in conjunction allow us to solve for both misalignment parameters, $\Delta i$ and $\Delta \Omega$. If spectral line profiles are also available, further improvement in the accuracy of these two parameters can generally be attained \citep{albrecht2007, philippov2013}.

The treatment of misalignment in PHOEBE 2.1 is warranted whenever the tidal and rotational distortion of misaligned stars are non-negligible. Depending on the precision of acquired data points and the degree of misalignment, this detailed treatment may or may not be warranted and spherical models might be adequate in terms of precision and superior in terms of computation time. We are not aware of any other public codes that deal with misalignment in deformed stars. The computational time cost is only marginally impacted by the addition of misalignment.



\subsubsection{The DI Herculis system}

\begin{figure}[!t]
\centering
\includegraphics[width=0.6\textwidth]{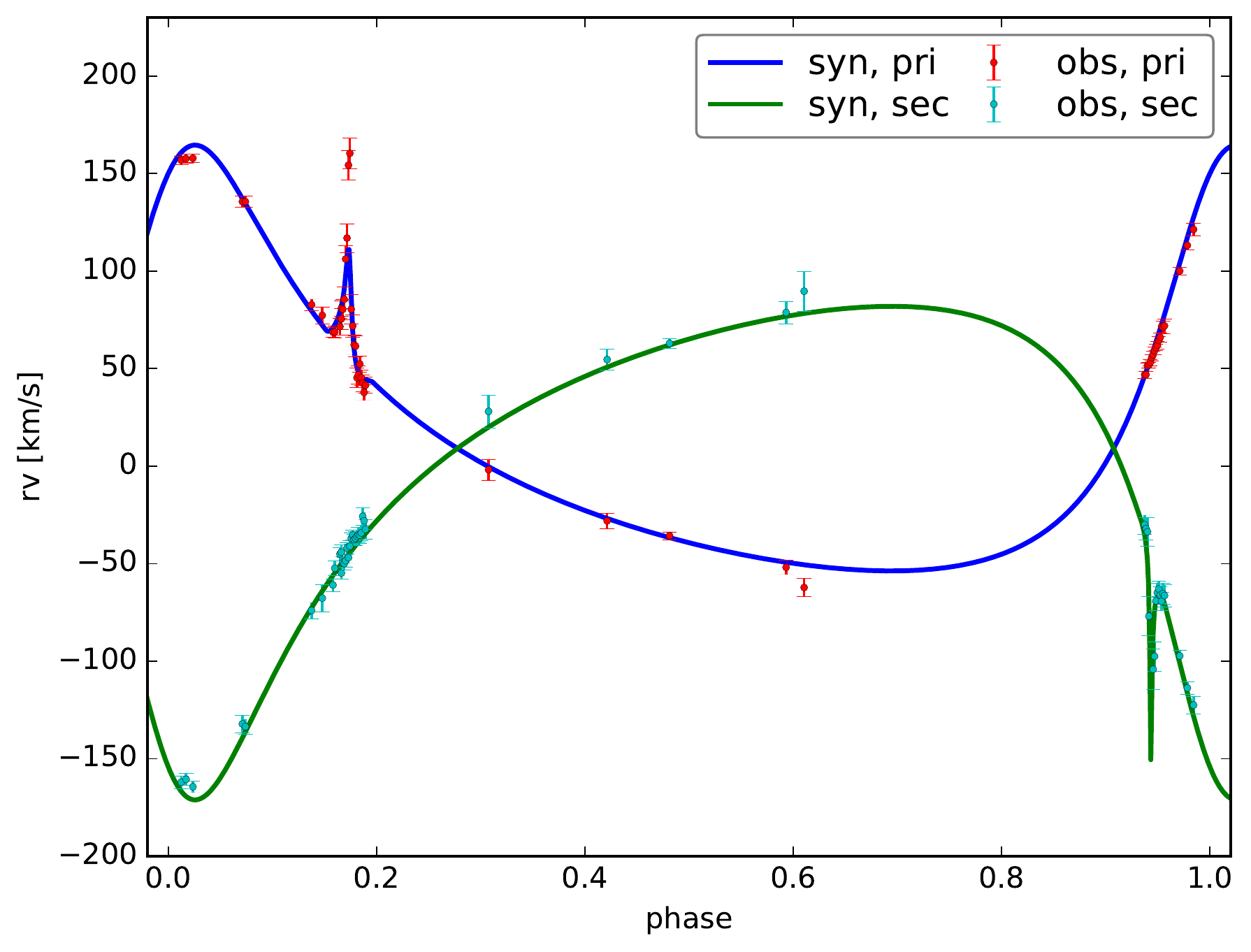}
\caption{Synthetic radial velocity curves of the primary (blue) and secondary star (green), plotted over the measured radial velocities from \citet{albrecht2009} for the primary (red) and secondary star (cyan) as the function of the phase $t/P$ with time $t$ measured from the primary eclipse.}
\label{fig:diher_rvs}
\end{figure}

\citet{albrecht2009} reported that DI Her is strongly misaligned, with the spin axes nearly perpendicular to the orbital axis. We use this example to further test and demonstrate the implementation of misaligned binary systems in \phoebe. Fig.~\ref{fig:diher_rvs} depicts the RV curves synthesized using the DI Her parameters summarized in Table \ref{tab:diher} plotted over the observed radial velocities from \citet{albrecht2009}. The misalignment parameters are taken from \citet{philippov2013}. We did not refit the data as that is beyond the scope of the current paper; we only report qualitative agreement with the published results.

\begin{table}[t]
\caption{
\label{tab:diher}
Principal parameters of the DI Herculis system.}
\begin{center}
\begin{tabular}{lccc}
  \hline \hline
  parameter: & & system: & \\
            & primary star: & & secondary star: \\ 
  \hline
  semi-major axis $a[R_\odot]$ & & 42.8731 & \\
  period $P[{\rm d}]$ & & 10.550164 & \\
  mass ratio $q$ & & 0.815 & \\
  eccentricity $\epsilon$ & & 0.489 & \\
  inclination $i [^\circ]$ & & 89.3 & \\
  long. of ascending node $\Omega[^\circ]$ & & 330.2 & \\
  systemic velocity $\gamma[{\rm km/s}]$ & & 9.1 & \\
  time of sup. conjunction $[{\rm d}]$ & & 2442233.3481 & \\
  \hline
  atmosphere & blackbody & & blackbody \\
  equivalent radius $R[R_\odot]$ &  2.68 & & 2.48\\
  effective temperature $T_{\rm eff}[K]$ &17300 & & 15400\\
  synchronicity parameter $F$ & 8.4819 & & 9.8487 \\
  mass $m[M_\odot]$ & 5.1 & &  4.4 \\
  \hline
  LD model & logarithmic & & logarithmic \\
  LD coefficient $x_{\rm LD}$ & 0.5 & & 0.5 \\ 
  LD coefficient $y_{\rm LD}$ & 0.5 & & 0.5 \\
  gravity darkening $\beta_\mathrm{grav}$ & 1 & & 1 \\
  \hline
  $\Delta \Omega[{}^\circ]$ & 72 & & -84  \\
  $\Delta i [{}^\circ]$ & 62 & & 100 \\
  \hline
\end{tabular}
\end{center}
\end{table}

\subsubsection{The Kepler-13Ab system}

Kepler-13Ab is a transiting hot Jupiter system with an A-type host star. It was first discovered as a misaligned system by \cite{szabo2011}, the only such system ever found without the accompanying Rossiter-McLaughlin affect. Since then the system has been widely studied, yet the models feature inconsistent parameter values. For this paper, we take a representative sample of these values, given in Table \ref{tab:kepler13}, and create a model light curve of this system.

\begin{table}[t]
\caption{
\label{tab:kepler13}
Principal parameters of the Kepler-13 System}
\begin{center}
\begin{tabular}{lccc}
  \hline \hline
  parameter: & & system: & \\
            & Kepler 13A : & & Kepler 13b: \\ 
  \hline
  semi-major axis $a[R_\odot]$ & & 7.36$^{\rm{a}}$ & \\
  period $P[{\rm d}]$ & & 1.763586522\footnote{\cite{muller2013}} & \\
  mass ratio $q$ & & 0.0037 & \\
  eccentricity $\epsilon$ & & 0.0 & \\
  inclination $i [^\circ]$ & & 85.82$^{\rm{a}}$ & \\
  \hline
  atmosphere & interpolated & & blackbody \\
  equivalent radius $R[R_\odot]$ &  1.69$^{\rm{a}}$ & & 0.144$^{\rm{b}}$ \\
  effective temperature $T_{\rm eff}[K]$ & 7650\footnote{\cite{shporer2014}} & & 2750$^{\rm{b}}$ \\
  synchronicity parameter $F$ & 0.591\footnote{\cite{szabo2014}} & & 1.0 \\
  mass $m[M_\odot]$ & 1.72$^{\rm{b}}$ & &  0.0063$^{\rm{b}}$ \\
  
  \hline
  LD model & interpolated & & logarithmic \\
  LD coefficient $x_{\rm LD}$ & -- & & 0.5 \\ 
  LD coefficient $y_{\rm LD}$ & -- & & 0.5 \\
  gravity darkening $\beta_\mathrm{grav}$ & 0.32 & & 0.32 \\
  \hline
  $\Delta \Omega[{}^\circ]$ & 4.52\footnote{\cite{masuda2015}} & & --  \\
  $\Delta i [{}^\circ]$ & 58.6\footnote{\cite{johnson2014}} & & -- \\
  \hline
\end{tabular}
\end{center}
\end{table}

\begin{figure}[!htb]
\centering
\includegraphics[width=15cm]{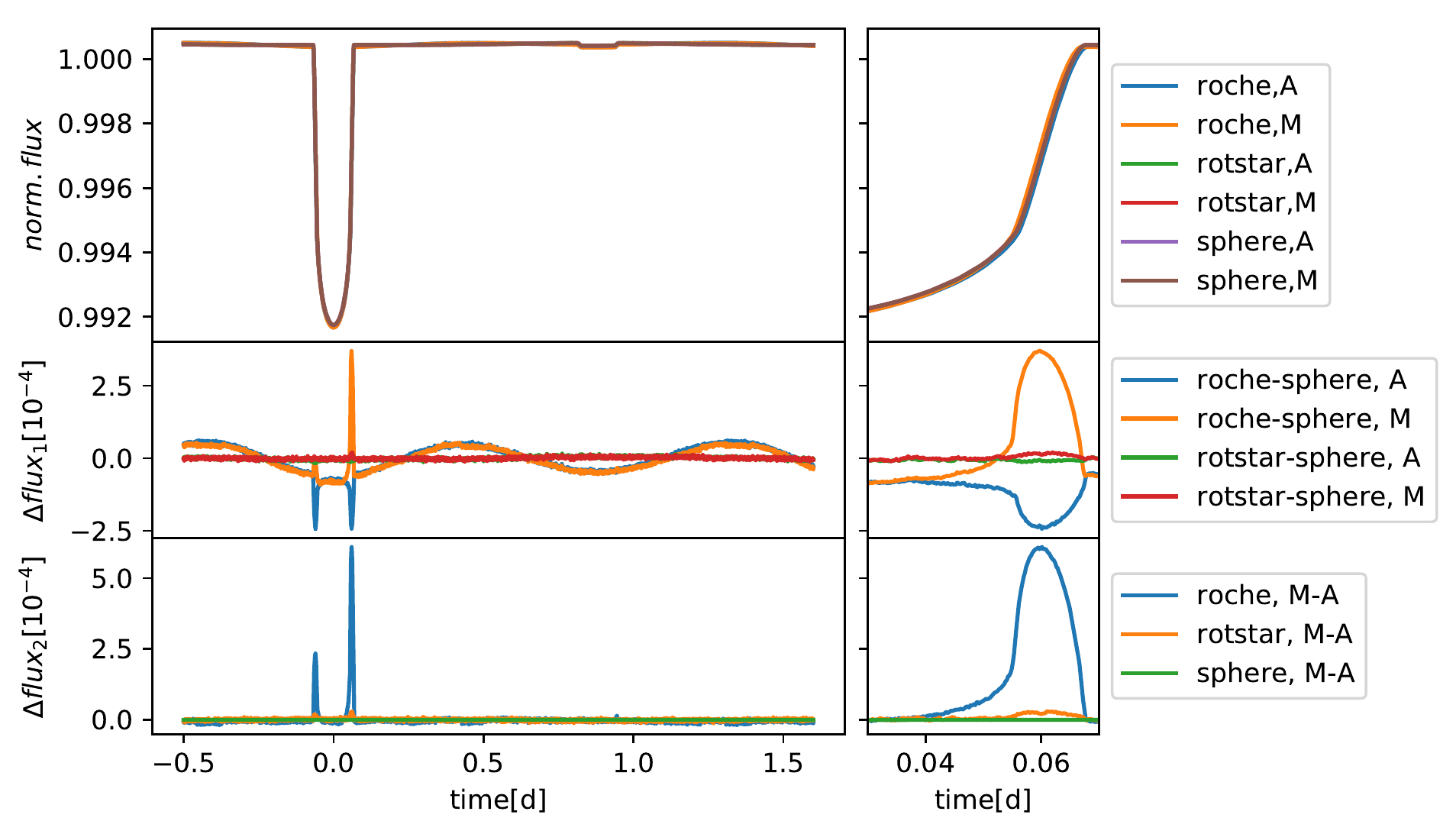}
\caption{Light curves normalized w.r.t. their integrals of the Kepler-13 system across the whole period (left panels) and a zoomed-in egress region of the primary eclipse (right panels) using different models. Upper panel: light curves that correspond to different models (Roche, sphere and rotating star), with $M$ and $A$ denoting misaligned and aligned realizations of the model. Middle panel: differences between model light curves and the corresponding spherical model. Bottom panel: differences between the misaligned and aligned realization of the models.}
\label{fig:kepler13}
\end{figure}

For comparison purposes we computed the light curves of the Kepler-13 system using different geometrical models: Roche, rotating star and spherical star. We kept the volume and misalignment of each star fixed between these different models. The light curves are normalized w.r.t. their corresponding integrals in order to make them more comparable to each other. The results are depicted in Fig.~\ref{fig:kepler13}. The differences between light curves are the largest in the ingress and egress of the primary eclipse. In the middle panel we compare light curves computed by different distortion models with the light curve obtained by the spherical model while keeping the volume and the degree of misalignment constant. We see that the largest discrepancy is in the Roche model, of the order of $\sim 3 \times 10^{-4}$, and it changes the signs depending on alignment. The bottom panel depicts the comparison of light curves of the misaligned model with the light curve of the aligned model, where we again see that the largest differences are in the Roche model, approximately equal to $\sim 6 \times 10^{-4}$. Thus, the effect in Kepler-13 is under 1 mmag, however that is well within \textsl{Kepler}'s precision reach of $\sim 20$-$30$\,ppm. Note that computing these differences accurately requires a sufficiently precise eclipsing algorithm, which is provided by \phoebe.

\section{Conclusion}

This paper summarizes the mathematical formalism of binary systems with the misaligned spin and orbital axes and introduces a new version of the modeling suite \phoebe that implements this formalism. The topic has been studied in the past, i.e.~by \citet{avni1982}, but to the best of our knowledge this is the first public implementation for the Roche-based geometry.



\begin{figure}[t]
\centering
\includegraphics[width=10cm]{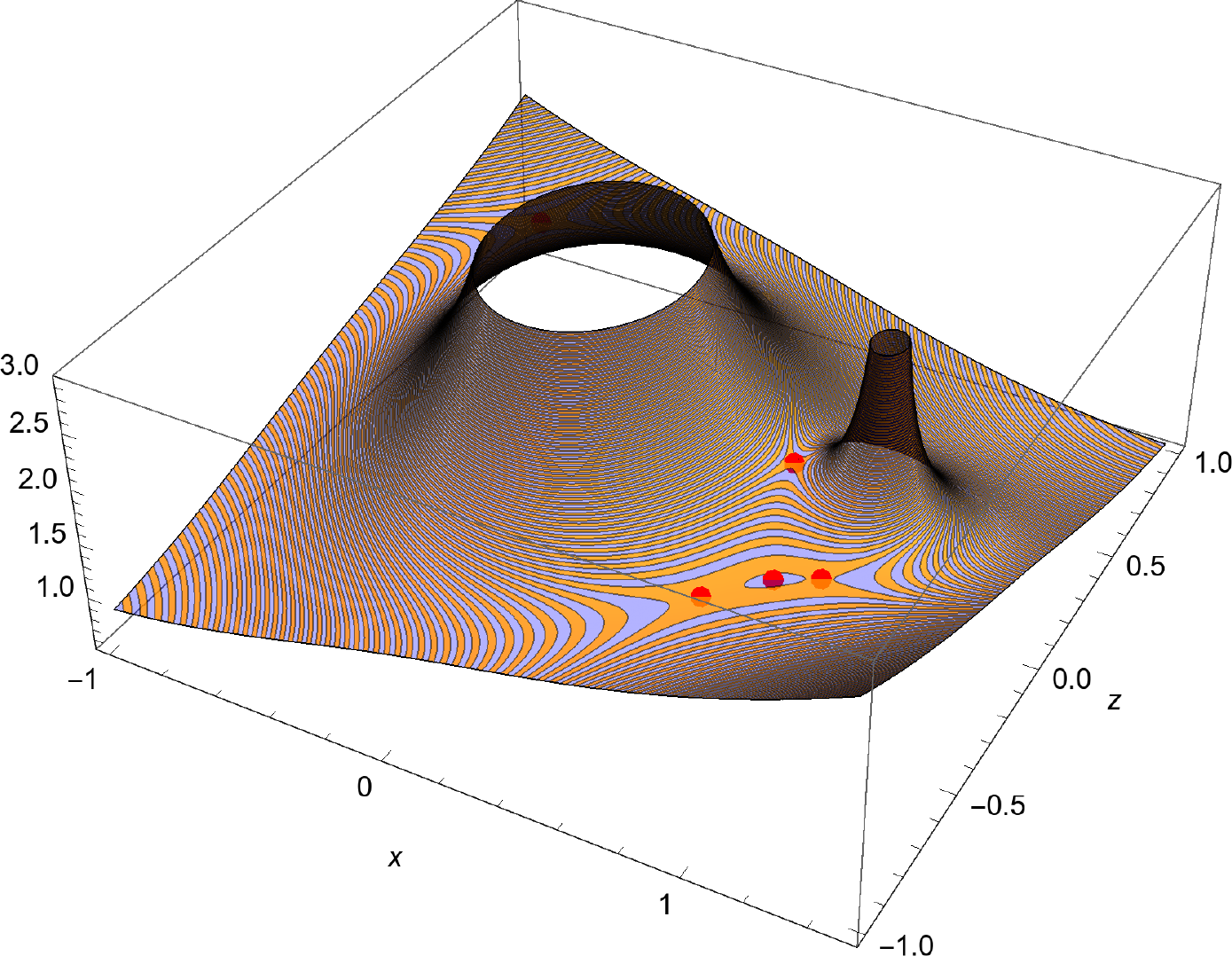}
\caption{A contour plot of the Kopal potential $\widetilde\Omega(x,0,z; q, b,\beta)$ (Eq. \ref{eq:poten_orig}) in the $xz$ plane at $q=0.1$ and  $b=(1+q) F^2 \delta^3$ for $F=1.1$, $\delta=1$ and $\beta=0.8$.}
\label{fig:pot_center}
\end{figure}

Beyond the anticipated systematic treatment of misalignment in eclipsing binary and extrasolar planet systems, a thorough study of the parameter space can yield some very interesting and readily testable predictions. For example, by varying the misalignment parameter $\beta$, we can find a local minimum near the lobes as depicted in Fig.~\ref{fig:pot_center}. Such islands of stability could harbor Trojan objects that are synchronized with the rotation of the primary star and reminiscent of the features seen in Tabby's star, KIC 8462852 \citep{boyajian2016}. Such hypotheses clearly merit further investigation beyond the scope of this introductory paper. The complexities of the nonlinear space spun by the parameters of misaligned objects predict many unexpected and intuition-challenging scenarios to exist in nature.

\acknowledgements
We would like to thank Simon Albrecht for his attentive review of the manuscript, pointing out several deficiencies and suggesting the discussion of photometric effects of misalignment in the case of Kepler-13. This work was supported by the NSF AAG grant \#1517474, which we gratefully acknowledge. A.P. and K.H. also acknowledge partial funding from the Slovenian Research Agency Grant P1-0188. K.H.~acknowledges the NASA ADAP grant 16-ADAP16-0201. KC is supported under NASA NESSF Fellowship \#NNX15AR87H. 

\bibliographystyle{aasjournal}
\bibliography{misaligned}

\appendix

\section{Derivation of the potential for a misaligned system} \label{sec:deriv_kopal}

A binary system consists of two stars, labeled $A$ and $B$. Their positions in the inertial (center-of-mass) coordinate system are denoted by $\vec{r}_\mathrm{A}$ and $\vec{r}_\mathrm{B}$. We assume that the center of mass of the binary system is at rest or moving with a constant velocity. Star A rotates as a rigid body about a misaligned axis $\uvec S$ with the angular velocity $\vec{\omega}_S$. The rigid body assumption asserts that every point on the primary star lobe co-rotates with the star. The equation of motion that describes the dynamics of the particle at position $\vec{r}$ is given by
$$
 \ddot{\vec{r}} = -\nabla U - \frac{1}{\rho}\nabla p \>,
$$
where $p$ is the pressure, $\rho$ is the particle density and $U$ is the gravitational potential of both stars:
$$
  U = 
  \frac{G M_\mathrm{A}}{\|\vec{r} - \vec{r}_\mathrm{A} \|} +
 \frac{G M_\mathrm{B}}{\|\vec{r} - \vec{r}_\mathrm{B} \|}.
$$
We now introduce a canonical coordinate system that is centered in star A, its $x$-axis points toward star B, its $z$-axis is aligned with the revolution axis $\vec L$, and it co-rotates with the center of star A in orbit about the common center of mass. We express vector $\vec{r}$ as the sum of the vector to the center of star A ($\vec r_\mathrm{A}$) and the vector relative to the center of star A ($\vec{r'}$):
$$
\vec r = \vec r_\mathrm{A} + \vec{r'}, \qquad {\vec{\ddot r}} = {\vec{\ddot r}_\mathrm{A}} +  {\vec{\ddot r'}} \>.
$$
The term $\vec{\ddot r}_\mathrm{A}$ describes the acceleration of star A caused by gravity:
$$
  \vec{\ddot r}_\mathrm{A} = G M_\mathrm{B} \frac{\vec{r_B}-\vec{r_A}}{\| \vec{r_B} - \vec{r_A} \|^3} \>.
$$
The term $\vec{\ddot r'}$ corresponds to the acceleration relative to the center of the primary star. To express it, we introduce a third coordinate system $S$ that co-rotates with the primary star itself about the rotation axis $\uvec S$. The relative vector $\vec{\ddot r'}$ is then
\begin{equation}    
  \vec{\ddot r'} =
 (\vec{\ddot r'})_S 
 + 2 \vec{\omega}_S \times ({\vec{\dot r'}})_S
 + \vec{\dot \omega}_S \times \vec{r'}
 + \vec{\omega}_S \times \vec{\omega}_S \times \vec{r'}\>.
\label{eq:rel_accel}
\end{equation}
We assume that the angular velocity is constant in time, so  ${\vec{\dot \omega}}_S=0$, and both the velocity and the acceleration in the co-rotating frame $S$ are zero: $(\vec{\ddot r'})_S = 0$, $(\vec{\dot r'})_S = 0$. The equation of motion thus takes the following form:
\begin{equation} \label{eq:motion}
 G M_\mathrm{B} \frac{\vec{r_B} - \vec{r_A}}{\| \vec{r_B} - \vec{r_A} \|^3} 
+ \vec{\omega}_S \times \vec{\omega}_S \times \vec{r'}
 = -\nabla U -\frac{1}{\rho} \nabla p\>.
\end{equation}
%

The first term in Eq.~(\ref{eq:motion}) can be written as the radial gradient $\nabla_{\vec{r'}}$ of the potential, where $\nabla_{\vec{r'}} \equiv \mathrm{d}/\mathrm{d}\vec{r'}$ operates in the canonical coordinate system spun by the basis vectors $(\ihat, \jhat, \khat)$:
$$
 G M_\mathrm{B} \frac{\vec{r_B} - \vec{r_A}}{\| \vec{r_B} - \vec{r_A} \|^3}
 \equiv G M_B \frac{\ihat}{d^2} =
 \nabla_{\vec{r'}} \left\{
  G M_\mathrm{B} \frac{\ihat \cdot \vec{r'}}{d^2} \right\},
$$
with $d$ being the distance between the stars, and similarly, by using the triple product rule $\vec a \times \vec b \times \vec c = \vec b (\vec a \cdot \vec c) - \vec c (\vec a \cdot \vec b)$, the last term in Eq.~(\ref{eq:rel_accel}) can be written as
$$
  \vec{\omega}_S \times \vec{\omega}_S \times \vec{r'} \equiv
  \omega_S^2
  \left[ (\uvec{S} \cdot \vec{r'}) \uvec S - \vec{r'} \right] =
  \nabla_{\vec{r'}} \left\{ \frac{1}{2} \omega_S^2
  \left[ (\uvec{S} \cdot \vec{r'}) \uvec S - \vec{r'} \right]^2 \right\}.
$$
Using these two expressions in the equation of motion yields
$$
  \nabla_{\vec{r'}} \left\{
  U + GM_\mathrm{B} \frac{\ihat \cdot \vec{r'}}{d^2}
  + \frac{1}{2}
  \left[ (\uvec{S} \cdot \vec{r'}) \uvec S - \vec{r'} \right]^2
  \right\}  = -\frac{1}{\rho}\nabla_{\vec{r'}} p.
$$
We can now readily recognize the expression within the curly braces as the (negative) potential $V$ of the misaligned binary system, as given in Eq.~(\ref{eq:kopal_potential}).

\section{Symmetries of the reduced Kopal potential}\label{sec:symm}

The reduced potential has several useful symmetries that have been implicitly used in the paper:
\begin{align*}
   \widetilde\Omega(x', y', z'; q, b, \beta) &=  
          \widetilde\Omega(x', -y', z'; q, b, \beta)\>,\\
   \widetilde\Omega(x', y', z'; q, b, \beta) &=  
          \widetilde\Omega(x', y', z'; q, b, \beta+\pi)\>,\\
  \widetilde\Omega(x', y', z'; q, b, \beta) &=  
           \widetilde\Omega(x', y', -z'; q, b, -\beta)\>.
\end{align*}

\section{Poles of misaligned lobes} \label{sec:poles}

The \emph{poles} are defined as the radii of the lobes $\lobe$ along the positive and negative direction of the spin vector $\uvec{S}$. In order to investigate this in more detail, we further rotate a coordinate system with the vector basis $(\ihat',\jhat',\khat')$, given by Eq.~(\ref{eq:rot_basis}), so that the new $z$-axis is aligned with the spin vector. The vector basis of this new coordinate system is:
\begin{equation}
        \ihat'' = \ihat' \cos\beta - \khat' \sin\beta\>,\qquad
        \jhat'' = \jhat'\>,\qquad
        \khat'' = \ihat' \sin\beta + \khat' \cos\beta\>,
    \label{eq:pole_basis}
\end{equation}
and position denoted by $\vec{\rho}'' = r (\sin\theta \cos\phi \, \ihat'' + \sin\theta\sin\phi \, \jhat'' + \cos\theta \, \khat'')$. In this coordinate system, the reduced Kopal potential can be rewritten as
\begin{equation}
  \begin{split}
    W(r,\theta,\phi; q, b,\beta) &=  \widetilde\Omega(\vec{\rho}''; q, b, \beta)\>, \\
    &= \frac{1}{r} + \frac{1}{2} b r^2 \sin ^2\theta - q r (\sin\beta\cos\theta+\cos\beta\sin\theta\cos\phi)\\
    & + \frac{q}{\sqrt{1-2 r (\sin\beta\cos\theta +\cos\beta \sin\theta\cos\phi)+r^2}}\>.
  \end{split}
  \label{eq:poten_red1}
\end{equation}
We see that $W$ does not have a quadratic term for the distance from the origin, which is associated with the centrifugal contribution to the potential. In general, the lobe is not symmetric across $z=0$ for the new coordinate frame and so the poles in the positive and negative directions of the rotating axis are not equal.

For a given set of parameters $(q,b,\beta)$ and the reference potential value $\widetilde \Omega_0$, the pole in the positive direction of the spin, $r_+$, and in the negative direction, $r_-$, are defined as
\begin{equation}
  W(r_+, 0, 0; q, b, \beta) =  W(r_-, \pi, 0; q, b, \beta) = \widetilde\Omega_0\>.    
\end{equation}
This yields the following equation for the poles:
\begin{equation}
    \frac{1}{r_\pm} +
    q \left(\frac{1}{\sqrt{1 \mp 2 \sin\beta r_\pm + r_\pm^2}} \mp \sin\beta r_\pm\right) = 
    \widetilde\Omega_0 \>.
\end{equation}
In the case of lobes with spin-orbit misalignment, it is more meaningful to discuss the diameter $r_+ + r_-$ along the rotation axis as a measure of the characteristic size of the object. These equations are solved in \phoebe by employing the standard Newton-Raphson method.

In the limit of large potential reference values, $\surfpot_0 \gg 1$, the poles can be approximated by a power series in $s = {\surfpot_0}^{-1}$, obtained by the inverse series method. The expansion of poles $r_\pm$ in the positive (+) and negative (-) direction are identical up to the 4th degree in $s$:
$$  
r_\pm = s + q s^2 + q^2 s^3 + \frac{1}{2} q s^4 (2q^2 + 3 \sin^2\beta - 1) +  O(s^5) \>,
$$
with the difference between the poles found only in terms of degree 5 and higher:
$$
    r_+ - r_- =  q (-3 + 5 \sin^2\beta) s^5 + O(s^6)\>.
$$
We see that, in the limit of a large potential, the lobe size depends only weakly on the misalignment parameter.

In general, the pole $r_\pm$ can be obtain by integrating the differential of the pole w.r.t.~the reciprocal potential $s$, given by:
$$
\frac{{\rm d} r_\pm}{{\rm d} s} = 
    \frac{g (g + q r_\pm \mp q g\sin\beta r_\pm^2)^2}
    {g^3 \pm q (g^3 - 1) \sin\beta r_\pm^2 + q r_\pm^3}\>,
$$
for $s \in [0,\widetilde \Omega_0^{-1}]$ and the initial condition $r_\pm(s=0) = 1$ and $g = \sqrt{1 \mp 2 \sin\beta r_\pm + r_\pm^2}$.

\section{Volume and area calculation}

We now turn our attention to the surface area ($\widetilde A$) and volume ($\widetilde V$) of the lobes $\lobe$ and the derivative of the volume w.r.t.~the value of the potential ($\widetilde V_{,\surfpot_0})$. We present a numerical method to compute these quantities using spherical coordinates $(r,\theta, \phi)$ and the reduced Kopal potential $W$ (Eq.\,\ref{eq:poten_red1}). We write a partial derivative of a function $f$ w.r.t. variable $x$ as $f_{,x} = \partial f/\partial x$. 

If $r(\theta,\phi)$ is known, the quantities are given by the following integrals:
\begin{align}    
  &\widetilde A = \int_0^\pi  \dot {\widetilde A}(\theta) \, {\rm d} \theta 
  &&\dot{\widetilde A}(\theta) = \frac{2}{3}
  \int_0^\pi r(\theta, \phi) \sqrt{r_{,\phi}^2 + \sin^2\theta(r^2 + r_{,\theta}^2)} \, {\rm d} \phi\>, \label{eq:A_def} \\
  &\widetilde V = \int_0^\pi \dot{\widetilde V} (\theta) \, {\rm d} \theta
  &&\dot{\widetilde V}(\theta) = \frac{2}{3} \sin\theta\int_0^\pi r^3(\theta, \phi) \, {\rm d} \phi\>, \label{eq:V_def} \\
  &\widetilde V_{,\widetilde \Omega_0} = \int_0^\pi \dot{\widetilde V}_{,\widetilde \Omega_0}(\theta) \, {\rm d} \theta 
  &&\dot{\widetilde V}_{,\widetilde \Omega_0}(\theta) = 
  2 \int_0^\pi \frac{r^2(\theta, \phi)}{W_{,r}(r(\theta,\phi), \theta, \phi)} \, {\rm d} \phi\>, \label{eq:dV_def}
\end{align}
where we took into account the symmetry over the $xz$ plane. The derivatives of the radius $r$ w.r.t.~spherical angles $(\theta,\phi)$ are given by
\begin{equation}    
  r_{,\theta}(r,\theta,\phi) = -\frac{W_{,\theta} (r,\theta,\phi)}{W_{,r}(r,\theta,\phi)}\>,
  \qquad
  r_{,\phi} (r,\theta,\phi) =  -\frac{W_{,\phi}(r,\theta,\phi)}{W_{,r}(r,\theta,\phi)} \>.
  \label{eq:r_deriv}
\end{equation}
The derivative $\widetilde V_{,\widetilde \Omega_0}$ is needed in the volume conservation process, whereby we find the value of the potential $\widetilde \Omega_0$ corresponding to a certain volume $\widetilde V_0$ as other parameters are fixed. This is analogous to calculating the inverse of $\surfpot_0 = {\widetilde V}^{-1} (\widetilde V_0)$ by the Newton-Raphson method: 
\begin{equation}
  \widetilde \Omega_{0, k+1} = 
  \widetilde \Omega_{0,k} - 
  \frac{\widetilde V(\widetilde \Omega_{0,k}) - \widetilde V_0}{V_{,\widetilde \Omega_0} (\widetilde \Omega_{0,k})}\>. 
  \label{eq:vol_cons} 
\end{equation}
%

We perform the calculation of $\widetilde A$, $\widetilde V$, and $\widetilde V_{,\widetilde \Omega_0}$ using two techniques: the integration across the surface and the asymptotic approximation in the limit of small lobes (large values of the potential). We explain both methods below.

\subsection{Integration over the surface}

The quantities $\widetilde A$, $\widetilde V$, and $\widetilde V_{,\surfpot_0}$ are written as definite integrals of their derivatives $\dot{\widetilde A}$, $\dot{\widetilde  V}$ and $\dot{\widetilde V}_{,\Omega_0}$ over the azimutal angle $\theta \in [0, \pi]$ per Eqs.~\ref{eq:A_def}, \ref{eq:V_def} and \ref{eq:dV_def}. The derivatives are given as integrals over the polar angles $\phi\in[0,\pi]$. We start the calculation by first approximating the derivatives and then integrate them across the total range of azimutal angle.

The integrals defining derivatives are calculated by discretizing the polar angle domain and using the Legendre-Gauss quadrature \citep{sirca2012}, whereby an integral of a function $g$ over the interval $[0,\pi]$ is approximated by:
\begin{equation}
  \int_0^\pi  g(\phi) \, {\rm d} \phi =
  \sum_{i=1}^n u_i g(\phi_i) + \frac{\pi^{2n+1}(n!)^4}{(2n+1)[(2n)!]^3}g^{(2n)}(\zeta) \>,
  \label{eq:lg_derived}
\end{equation}
where $u_i$ and $\phi_i$ are appropriately chosen weights and nodes, respectively, and $\zeta \in [0,\pi]$. The weights and nodes are given by
$$
  u_i = \frac{\pi}{2} w_i\>,\qquad \phi_i =\frac{\pi}{2} (1+x_i) \>,
$$
where $w_i$ and $x_i$ are standard Legendre-Gauss weights and nodes, respectively, determined for functions integrated over range $[-1,1]$.

The radius $r(\theta,\phi)$ of the lobe at arbitrary angles $\theta$ and $\phi$ can be obtained by integrating
\begin{equation}
    \frac{{\rm d} r }{{\rm d} \theta} \equiv r_{,\theta} (r, \theta, \phi_i) \>,
    \label{eq:int_surf}
\end{equation}
with the initial condition $r(\theta=0,\phi) = r_+$, where the derivative $r_{,\theta}$ is given by Eq.~(\ref{eq:r_deriv}). For each polar angle $\phi_i$ we introduce a radius $r_i(\theta) = r(\theta,\phi_i)$ along the azimuthal angle $\theta$. By using Legendre-Gauss quadrature (Eq.~\ref{eq:lg_derived}) we approximate the integrals defining derivatives $\dot{\widetilde A}$, $\dot{\widetilde V}$ and $\dot{\widetilde V}_{,\widetilde \Omega_0}$ as sums over the set of functions $\{r_i\}_{i=1}^n$. Then, by taking into account Eq.~(\ref{eq:int_surf}), we rewrite $\widetilde A$, $\widetilde V$ and ${\widetilde V}_{,\widetilde \Omega_0}$ as a solution to $n+3$ ordinary differential equations:
\begin{align}
    &\frac{{\rm d} r_i }{{\rm d} \theta } 
        = r_{,\theta} (r_i(\theta), \theta, \phi_i) \qquad i = 1,\ldots, n \>,& \\  
    &\frac{{\rm d} \widetilde A }{{\rm d} \theta} 
        = 2 \sum_{i=1}^n u_i r_i(\theta) 
        \sqrt{r_{,\phi}^2(r_i(\theta),\theta,\phi_i) + \sin^2\theta \left(r_i^2(\theta) + r_{,\theta}^2(r_i(\theta),\theta,\phi_i)\right)} \>,& \\
    &\frac{{\rm d} \widetilde V }{{\rm d} \theta} 
        = \frac{2}{3} \sin\theta \sum_{i=1}^n u_i r_i^3(\theta) \>,&\\  
    &\frac{{\rm d}\widetilde V_{,\widetilde \Omega_0} }{{\rm d} \theta} 
        = 2 \sum_{i=1}^n u_i \frac{r_i^2(\theta)}{W_{,r}(r_i(\theta),\theta,\phi_i)} \>,&
\end{align}
which are integrated for $\theta \in [0,\pi]$ with the initial conditions
\begin{align}
    &r_i(\theta = 0) = r_+  \qquad\qquad  i = 1,\ldots, n\>,\\
    &\widetilde A(\theta=0) = 
    \widetilde V(\theta=0) = 
    \widetilde V_{,\widetilde \Omega_0}(\theta=0) = 0\>. 
\end{align}
The quantities in question are then obtained at $\theta=\pi$:
$$
 \widetilde A = \widetilde A(\theta=\pi),\qquad 
 \widetilde V = \widetilde V(\theta=\pi),\qquad
 \widetilde V_{,\widetilde\Omega_0} = \widetilde V_{,\widetilde \Omega_0}(\theta=\pi)\>.
$$
Note that $r_i(\theta=\pi) = r_-$, which can be used as a numerical check of integration.

\subsection{The limit of small lobes}

In the limit of large $\surfpot_0$, the radius can be expressed as a power series of $s ={\widetilde\Omega_0}^{-1}$ using the inverse series technique, which we can symbolically write as
\begin{equation}
    r(\theta,\phi; q, b, s,\beta) = \sum_{k=1}^\infty a_k(\theta,\phi;q,b,\beta) s^k \>.
    \label{eq:r_ser}
\end{equation}
By plugging Eq.~(\ref{eq:r_ser}) into the formul\ae\ for area $\widetilde A$ (Eq.\,\ref{eq:A_def}) and volume $\widetilde V$ (Eq.\,\ref{eq:V_def}), we obtain their own expansions in $s$ and write them as
\begin{equation}
    \widetilde A = \frac{4 \pi}{{\widetilde \Omega_0}^2} {\cal A}(s;q,b,\beta) \qquad
    \widetilde V = \frac{4 \pi}{3 \widetilde\Omega_0^3} {\cal V} (s;q,b,\beta) \>, 
    \label{eq:AV_asymp}
\end{equation}
with auxiliary expressions
\begin{align}
    \begin{split}
     {\cal A}(s;q,b,\beta) 
      &= 1+2 q s+3 q^2 s^2+s^3 \left(\frac{2 b}{3}+4 q^3\right)+ 5 s^4 \left(\frac{2 b q}{3}+q^4\right) + s^5 \left(10 b q^2+6 q^5\right)\\
      & +s^6 \left(b^2+\frac{70 b q^3 - bq}{3}+b q \cos (2\beta) +7 q^6+2 q^2\right) + \ldots\>,
    \end{split} \label{eq:A_ser} \\
    \begin{split}
     {\cal V}(s;q,b,\beta) 
      &= 1+3 q s+6 q^2 s^2+s^3 \left(b+10 q^3\right)+s^4 \left(6 b q+15 q^4\right)+21 s^5 \left(b q^2+q^5\right)\\
      & + \frac{2}{5} s^6 \left(4 b^2+140 b q^3+3 b q \cos (2\beta)-b q+70 q^6+6 q^2\right)+ \ldots\>.
    \end{split} \label{eq:V_ser}
\end{align}
The influence of misalignment is thus very weak in the limit of small lobes and it affects only the terms of the 6th degree and higher in the series expansion. 

The expression for $\widetilde V_{, \widetilde\Omega_0}$, needed in the volume conservation procedure defined by Eq.~(\ref{eq:vol_cons}), is obtained by taking the derivative of the volume $\widetilde V$ (Eq.~\ref{eq:AV_asymp}) w.r.t. the reference potential value $\surfpot_0$:  
\begin{equation}
    \widetilde V_{,\widetilde \Omega_0} = 
    -\frac{4\pi}{3 {\widetilde \Omega_0}^4} 
    \left(3{\cal V}(s; q,b,\beta) + s \frac{{\rm d} {\cal V}(s; q,b,\beta)}{{\rm d} s}\right) \>.  
\end{equation}

\section{Rotation matrices}\label{sec:rot_mat}

In the paper we use the following convention for the rotation matrices in three dimensions with rotation angle $\phi$:
$$
  {\bf R}_u(\phi) = 
  \begin{bmatrix}
  1 & 0 & 0 \\
  0 & \cos\phi & -\sin\phi \\
  0 & \sin\phi & \cos\phi 
  \end{bmatrix}
\quad
  {\bf R}_v(\phi) = 
  \begin{bmatrix}
  \cos\phi & 0 & \sin\phi \\
  0 & 1 & 0 \\
  -\sin\phi & 0 & \cos\phi 
  \end{bmatrix}
\quad
  {\bf R}_w(\phi) = 
  \begin{bmatrix}
  \cos\phi & -\sin\phi & 0 \\
  \sin\phi & \cos\phi & 0 \\
  0 & 0 & 1
  \end{bmatrix}
$$

\end{document}